\documentclass{elsart}
\usepackage{graphicx,color,epsfig}
\begin{document}
\begin{frontmatter}
\title{Description of the $^{17}$F(p,$\gamma$)$^{18}$Ne radiative capture reaction in the continuum shell model}
\author[a]{R. Chatterjee,}
\author[b]{J. Oko{\l}owicz,}
\author[a]{and M. P{\l}oszajczak}

\address[a]{Grand Acc\'el\'erateur National d'Ions Lourds (GANIL),
CEA/DSM-CNRS/IN2P3, BP 55027, F-14076 Caen Cedex 05, France}
\address[b]{Institute of Nuclear Physics, Radzikowskiego 152, PL-31342 Krak\'ow, 
Poland}

                                                                                
\begin{abstract}
The shell model embedded in the continuum is applied to calculate the astrophysical S-factor and the reaction rate for the radiative proton capture reaction $^{17}$F(p,$\gamma$)$^{18}$Ne. 
The dominant contribution to the cross-section at very low energies is due to M1 transitions $J_i^{\pi}=2^+\rightarrow J_f^{\pi}=2_1^+$ whose magnitude is controlled by a weakly bound $2_2^+$ state at the excitation energy $E_x=3.62$ MeV. 
 \end{abstract}
\end{frontmatter}


\section{Introduction}

Radiative capture reactions play a significant role in nuclear astrophysics
being one of the important processes in stellar nucleosynthesis \cite{bur,rol1}.
Nova explosions in binary star systems occurring due to mass transfer
to a white dwarf star from its companion is an important site for the CNO
cycle. A thermonuclear runway results from the accreted matter accumulated
on the outer layers of white dwarf star, during which $^{17}$F could
be produced during hydrogen burning of the CNO elements. At this
juncture the $^{17}$F(p,$\gamma$)$^{18}$Ne radiative capture reaction  
becomes very important in deciding the fate of $^{17}$F and the subsequent
routes of nucleosynthesis. The proton capture rate of $^{17}$F is also
important because in competition with the 
$^{17}$F-$\beta$-decay rate it can provide an alternate path from the hot
CNO cycle to the rapid proton (rp) burning process via the 
$^{14}$O($\alpha$,p)$^{17}$F(p,$\gamma$)$^{18}$Ne($\beta^+,\nu$)
$^{18}$F(p,$\gamma$)$^{19}$Ne reaction sequence.
A smaller $^{17}$F(p,$\gamma$)$^{18}$Ne reaction rate compared to 
the $^{17}$F-$\beta$-decay rate at characteristic nova temperature
and densities would favour $^{15}$O enrichment via the reaction sequence
$^{17}$F($\beta^+,\nu$)$^{17}$O(p,$\alpha$)$^{14}$N(p,$\gamma$)$^{15}$O, 
whose $\beta$-decay to $^{15}$N could be an explanation for the overabundance of $^{15}$N
in nova ejecta. On the other hand a large $^{17}$F(p,$\gamma$)$^{18}$Ne reaction rate would alter the 
$^{18}$F/$^{17}$F abundance ratio due to the activation of the 
$^{17}$F(p,$\gamma$)$^{18}$Ne($\beta^+,\nu$)$^{18}$F reaction chain.
A good knowledge of the $^{17}$F(p,$\gamma$)$^{18}$Ne reaction rate
is also thought to be important for understanding the astrophysical
phenomenon of x-ray bursts.

It was first predicted in Ref. \cite{wish} that the $3^+$, strong $s$-wave resonance in $^{18}$Ne
would occur at an excitation energy $E_x = 4.328$ MeV with the  width $\Gamma = 5$ keV. 
This low energy $3^+$ was expected  to substantially contribute to the 
$^{17}$F(p,$\gamma$)$^{18}$Ne reaction rate above stellar temperatures greater
than 0.2 GK, which would be important for nova explosions. Since then there
have been several attempts \cite{garcia,hahn,park} to experimentally 
measure the exact position of the $3^+$ level in $^{18}$Ne and it was only recently \cite{bard1,bard}
that this has been unambiguously put at an excitation energy $E_x = 4.524$ MeV (center of mass
energy (c.m.) $E_r = 600$ keV) and width $\Gamma = 18$ keV. 

The direct capture part of the $^{17}$F(p,$\gamma$)$^{18}$Ne reaction has been 
studied largely in a potential model framework following mainly the work of Rolfs \cite{rolfs}. 
A more microscopic approach was the extended two-cluster model (ETCM) \cite{des}, 
using the generator coordinate method, which has also been applied to 
neutron rich Carbon isotopes \cite{des_car}. One of the main advantages of this method is that the resonant and non-resonant part of the reaction need not be calculated separately. 
However the configuration space which is used to describe the many-body states is
restricted to a few cluster configurations.

Weakly bound states or resonances cannot be described in the closed
quantum system (CQS) formalism  \cite{Dob98}. For bound states, there appears
a virtual scattering into the continuum phase space  involving
intermediate scattering states. Continuum coupling of this kind affects
also the effective nucleon-nucleon interaction. For unbound states,
the continuum structure appears explicitly in the properties of those states. 

The importance   of the particle continuum was discussed in the early days of
the multiconfigurational Shell Model (SM). However, a unified description of nuclear structure and nuclear reaction aspects became possible in
realistic situations only recently in the framework of the Shell Model Embedded in the Continuum (SMEC) \cite{karim,phyrep} which considers the nucleus as a quantum many-body system coupled to the environment of decay channels (the open quantum system (OQS) formalism).  In the SMEC, all couplings involving discrete states and scattering states are calculated using the SM interaction. Inclusion of the coupling between discrete and scattering states leads to a non-hermiticity of a many-body Hamiltonian that consists of the closed system with discrete eigenstates (the standard SM Hamiltonian), and the coupling  between the closed system Hamiltonian and its environment of one-nucleon \cite{karim} and two-nucleon \cite{jim1,rop} decay channels. Above the particle-emission threshold, the eigenvalues of the OQS are complex. Different continuum shell-model approaches \cite{csm1,csm2,csm3,volya}, including the SMEC \cite{karim,phyrep,rop}, are formulated in the Hilbert space, i.e., they are based on the completeness of a single-particle (s.p.) basis consisting of bound orbits and a real continuum.  A different approach to the treatment of particle continuum is proposed in the Gamow Shell Model \cite{Mic02}, which is the multi-configurational SM with  a s.p.~basis given by the Berggren ensemble \cite{Ber68} consisting of Gamow (or resonant) states and the complex non-resonant continuum of scattering states. Gamow states and the GSM can be formulated rigorously in the rigged Hilbert space (see Ref. \cite{Rig1} and references quoted therein).

In this work we present a study of the $^{17}$F(p,$\gamma$)$^{18}$Ne
radiative capture reaction using the formalism of SMEC \cite{karim,phyrep}. This formalism
has been applied before to describe the structure of mirror nuclei $^{8}$B,$^{8}$Li, 
and capture cross sections for mirror reactions  $^{7}$Be(p,$\gamma$)$^{8}$B, $^{7}$Li(n,$\gamma$)$^{8}$Li
\cite{karim,karim0} as also to the description of $^{16}$O(p,$\gamma$)$^{17}$F and 
$^{16}$O(p,p)$^{16}$O reactions \cite{nic1,nic2}.
The paper is organized in the following way. In section {\bf 2}, we present
a brief description of the SMEC formalism emphasising those aspects of the SMEC which
imply the modification of the CQS eigenvalues and eigenfunctions 
and attempts to put the theory of nuclear structure and reactions on the same footing. We also indicate and give the relevant formulae to calculate radiative capture cross sections
from both the ground state (g.s.) and excited states of the target nucleus, with properly
antisymmetrized wave functions being used in both initial and final states.
Section {\bf 3} contains the results and discussions of SMEC calculations
of the radiative capture reaction $^{17}$F(p,$\gamma$)$^{18}$Ne. The
self-consistent potential, entering the coupled-channel (CC) equations of SMEC
, and the parameters of the density-dependent residual interaction are presented in section {\bf 3.1}.
The spectrum of $^{18}$Ne obtained both in the SM and the SMEC  is in section {\bf 3.2}. The next two subsections {\bf 3.3} and {\bf 3.4} contain the astrophysical $S$-factors and the 
reaction rates of the $^{17}$F(p,$\gamma$)$^{18}$Ne reaction, respectively.
 Finally the summary and conclusions are presented in section  {\bf 4}.

\section{Formalism}
\subsection{Shell Model Embedded in the Continuum}
A unified description of interdependent nuclear
structure and nuclear reactions is attempted in SMEC  by using a projection operator technique \cite{Fesh}. For the purpose of describing the one-nucleon capture reactions, we assume that  the Hilbert space is decomposed
into two subspaces $Q$ and $P$:
\begin{eqnarray}
P + Q = I_d
\end{eqnarray}
The $Q$ subspace consists of $A$ particles  
in many-body localized states which are built up by the bound state s.p.
wave functions and the part of s.p. resonance wave functions localized inside the nucleus, i.e. before
a cutoff radius $R_{cut}$. The $P$ subspace contains $(A-1)$-particle states
in localized states built up from (quasi-)bound single particle (qbsp) orbits and one nucleon
in the scattering state. The rest of the s.p. resonant wave function (the
$Q$ subspace contains the part $r < R_{cut}$ of the resonance wave function)
outside the cutoff radius $R_{cut}$ are included in the $P$ subspace. The wave
functions in $Q$ and $P$ are then renormalized in order to 
ensure their orthogonality in both subspaces. The fully antisymmetrized
wave function can then be used to study the radiative capture processes 
 where the capture can occur from both the g.s. and the excited state of the
target nucleus $A-1$. This method is also fully symmetric in treating the
resonant and non-resonant part of the reaction and one does not need to
calculate them separately.

The first step in SMEC involves the generation  of localized states in the 
$Q$ subspace without any coupling to the continuum. This involves solving
the standard eigenvalue problem:
\begin{eqnarray}
H_{QQ} \Phi_i = E_i \Phi_i
\end{eqnarray}
where $H_{QQ} \equiv QHQ$ is identified with the SM Hamiltonian and $\Phi_i$
are the localized many-body wave functions with eigenenergies $E_i$.

The second step consists  of solving the CC equations:
\begin{eqnarray}
\label{homcc}
(E^{(+)} - H_{PP})\xi_E^{c(+)} = 0
\end{eqnarray}
where $H_{PP} = PHP$ and $'+'$ stands for the outgoing boundary conditions. 
The index $c \equiv (I_t,l,s,j)$ denotes different  channels characterized by $I_t$, the spin 
of the $(A-1)$ system, 
$s$, $l$ and $j$ which are the spin, orbital angular momentum and the total angular momentum, respectively, of the particle in the continuum. In a more explicit manner, eq. (\ref{homcc}) is written as \cite{phyrep}:
\begin{eqnarray}
\label{esp1}
\sum_{c^{'}}^{}(E^{(+)} - H_{cc^{'}}) {\xi}_E^{c^{'}(+)} = 0 
\end{eqnarray}
where
\begin{eqnarray}
\label{esp2}
H_{cc^{'}} = (T_{kin} + U ){\delta}_{cc^{'}} + {\upsilon }_{cc^{'}}^{J} 
\end{eqnarray}
In the above equation, $T_{kin}$ stands for the kinetic-energy operator and $U$ is the finite-depth potential of Woods-Saxon (WS) type with the spin-orbit term:
\begin{eqnarray}
\label{pot}
U(r) = V_0f(r) + V_{SO} (4{\bf l}\cdot{\bf s})
\frac{1}{r}\frac{df(r)}{dr}  + V_C  
\end{eqnarray}
where
\begin{eqnarray}
\label{pot1}
f(r) = \left[ 1 + \exp \left( \frac{r-R_0}{a} \right) \right]^{-1} 
\end{eqnarray}
with the Coulomb potential $V_C$  calculated for a uniformly charged sphere of radius $R_0$. 
$ {\upsilon }_{cc^{'}}^{J}$ in (\ref{esp2}) is the channel-channel coupling generated by the
residual interaction 
between the  $Q$ and $P$ subspaces and depends on
the target states of the $(A-1)$ system, s.p. orbitals $l,j$ 
and total angular momentum and parity $J^\pi$ of the $A$-particle system. The explicit formulae for $H_{cc^{'}}$ and $ {\upsilon }_{cc^{'}}^{J}$ is given in Ref. \cite{karim}.
As a residual coupling between the $Q$ and $P$ subspaces, we take the density-dependent contact force \cite{wam}:
\begin{eqnarray}
\label{force}
H_{PQ} = \nu\{ \rho(r) {\hat v}_{00}^{in} &+& [1 - \rho(r)]{\hat v}_{00}^{ex}
\nonumber \\
       &+& {\bf \tau}_1.{\bf \tau}_2[\rho(r) {\hat v}_{01}^{in} + 
                [1 - \rho(r)]{\hat v}_{01}^{ex}] \} \delta ({\bf r}_1-{\bf r}_2)
\end{eqnarray}
where $\nu$ is the relative strength parameter and $\rho(r)$ is the nuclear 
density having a Fermi form:
\begin{eqnarray}
\label{den}
\rho(r) = \left[ 1 + \exp \left( \frac{r-r_n}{d} \right) \right]^{-1}
\end{eqnarray}
with $r_n$ and $d$ defining the size and surface thickness of the nucleus, 
respectively. The values taken for the parameters are mentioned in 
the next section.

The diagonal part of ${\upsilon }_{cc^{'}}^{J}$ changes
the average potential $U$ and consequently the s.p. wave functions.
So a self-consistent procedure is adopted to calculate the wave functions
of those s.p. states which are occupied in the target nucleus. This iterative
procedure yields a new self-consistent average potential:
\begin{eqnarray}
\label{usc}
U^{(sc)}(r) = U(r)+{\upsilon }_{cc}^{J(sc)}(r) 
\end{eqnarray}
which is used to solve eq. (\ref{esp1}).

The third system of equations:
\begin{eqnarray}
\label{coup}
(E^{(+)} - H_{PP}){\omega}_{i}^{(+)} = H_{PQ}{\Phi}_i \equiv w_i
\end{eqnarray}
defines the functions ${\omega}_{i}$ that describe the continuation of the SM states $\Phi_i$
into the continuum. The source term $w_i$ couples the wave function of the $Q$ subspace with
those of the $P$ subspace.

Eigenstates of the CQS coupled to the external environment of decay channels are found by
solving the eigenvalue problem for the effective Hamiltonian:
\begin{eqnarray}
\label{heff}
H_{QQ}^{eff}(E) = H_{QQ} + H_{QP}G_{P}^{(+)}(E)H_{PQ} 
\end{eqnarray}
in the function space ($Q$) of the discrete states. $G_{P}^{(+)}(E)$~ is the 
Green function for the motion of a s.p. in the $P$ subspace and is given by:
$$G_P = P(E - H_{PP})^{-1} P~ \ .$$ 
The SM Hamiltonian $H_{QQ}$  is hermitian. On the contrary, $H_{QQ}^{eff}$ is an energy dependent, 
non-hermitian (complex and symmetric matrix) operator above the particle emission threshold
 and a hermitian (real) operator below the emission threshold.  The $Q-P$ coupling contained in $H_{QQ}^{eff}$ introduce an {\em external} mixing of SM states (eigenstates of $H_{QQ}$) via the coupling to the decay channels.  Diagonalization of $H_{QQ}^{eff}$
by the orthogonal but in general non-unitary transformation:
\begin{eqnarray}
\label{transf}
{\Phi}_i \longrightarrow {\tilde {\Phi}_j} = {\sum}_{i}^{} b_{ji}{\Phi}_i 
\end{eqnarray}
yields complex eigenvalues ${\tilde {E_i}} - \frac{1}{2}i{\tilde {{\Gamma}_i} }$ ,
which depend on the energy $E$ of the particle in the continuum.
The coefficients $b_{ji}$ in (\ref{transf}) form a complex matrix of eigenvectors in the SM basis satisfying
\begin{eqnarray}
\label{trf2}
\sum_k b_{jk}b_{ik} = \delta_{ji}
\end{eqnarray}
The eigenvalues of $H_{QQ}^{eff}(E)$~ at energies  $\tilde{E_i}(E) = E$,
determine the energies and widths, ${\tilde {{\Gamma}_i} }$, of resonance states.

The total wave function, ${\Psi}_{E}^{c(c_0)}$, for the many-body problem is 
now obtained in terms of ${\Phi}_i$~, ${\xi}_{E}^{c}$~, ${\omega}_i$ and the eigenvalues of
$H_{QQ}^{eff}$ as:
\begin{eqnarray}
\label{wf}
{\Psi}_{E}^{c(c_0)} = {\xi}_{E}^{c(c_0)} + \sum_{i}^{}{\tilde {\Omega}_i}
\frac{1}{E - {\tilde E_i} + (i/2){\tilde {\Gamma}_i}} 
\langle{\tilde {\Phi}_i} \mid H \mid {\xi}_{E}^{c(c_0)}\rangle
\end{eqnarray}
with the incoming wave only in channel $c_0$ as indicated by the superscript, and
\begin{eqnarray}
{\tilde {\Omega}_i} = {\tilde {\Phi}_i} + {\tilde {\omega}_i}
                    = (1 + G_{P}^{(+)}H_{PQ}){\tilde {\Phi}_i}
\end{eqnarray}
with ${\tilde {\omega}_i}$ defined by $G_{P}^{(+)}H_{PQ}{\tilde {\Phi}_i}$.
It is to be noted here that the SMEC formalism is fully symmetric in treating
the continuum and bound state parts of the solution: ${\Psi}_{E}^{c(c_0)}$
represents the continuum state modified by the discrete states and 
${{\tilde \Omega}}_i$ represents the discrete state modified by coupling to the embedding
continuum states.

The asymptotic conditions of Eq. (\ref{wf}) have been analyzed in Ref. \cite{csm1}.
One can then calculate the amplitude of the partial decay width  as:
\begin{eqnarray}
\label{pw}
{{\tilde \gamma}_i^c} = {\sqrt{2\pi}} \left({{4m_r}\over {\hbar^2k_c}}\right)^{1/2}
  \sum_j  b_{ji}\sum_{c'}\int^\infty_0 dr {\xi}_{E}^{c'(c)}(r) w_j^{c'}(r)
\end{eqnarray}
Using the proportionality  relation between the matrix elements in Eq. (\ref{wf})
and the amplitudes of the partial widths (eq. (\ref{pw})), one can derive the  scattering matrix ($S$-matrix):
\begin{eqnarray}
S_{c(c_0)} = S_{c(c_0)}^0 - i\sum_j
\frac{{{\tilde \gamma}_j^c} {{\tilde \gamma}_j^{c_0}} }
            {E - {\tilde E_j} + (i/2){\tilde {\Gamma}_j}} 
\end{eqnarray}
In the above equation $S_{c(c_0)}^0$ is determined from the
asymptotic, large distance behaviour of ${\xi}_{E}^{c(c_0)}$.
The amplitudes of the partial widths ${{\tilde \gamma}_j^c}$ as well as the
complex eigenvalues ${\tilde E_j} - (i/2){\tilde {\Gamma}_j}$ of the
effective Hamiltonian $H_{QQ}^{eff}$ which enter in the $S$-matrix,
are energy dependent.  ${{\tilde \gamma}_j^c}$ are also complex
due to the channel-channel coupling and the external mixing of SM eigenstates via the coupling to the continuum. The partial width for channel $c$ can then be defined by
$\Gamma_j^c = \mid \gamma_j^c\mid^2$, though the total width is no longer a
sum of partial widths \cite{mahaux}:
\begin{eqnarray}
\sum_c \mid \gamma_j^c\mid^2 = {{\tilde \Gamma}_j} \sum_j\mid b_{ji}\mid^2
\geq {{\tilde \Gamma}_j}
\end{eqnarray}
This is because the complex eigenvectors $b_{ji}$ are normalized in the sense of
eq. (\ref{trf2}), which implies:
\begin{eqnarray}
\sum_j\mid b_{ji}\mid^2 \geq 1
\end{eqnarray}
and which again is a direct consequence of the non-hermitian nature 
of the effective Hamiltonian $H_{QQ}^{eff}$ in $Q$.

Thus the formalism of SMEC provides a way to describe nuclear structure
and reactions in a single framework starting from the same Hamiltonian.
In the next sub-section we shall use the total wave function (\ref{wf}) to describe the
radial part of the initial state wave function in a radiative capture process.
More details about SMEC with one particle in the continuum, and general
discussions about the OQS could be found in Ref. \cite{phyrep}.

\subsection{Radiative capture}
We now give the general formulae to calculate radiative capture cross sections
from both the g.s. and excited states of the target nucleus, with 
antisymmetrized wave functions used in both initial and final states. We shall apply this to the reaction $^{17}$F(p,$\gamma$)$^{18}$Ne, where the proton capture can occur from both
$I_t=5/2^+$ g.s. and $I_t=1/2^+$ first excited state of the target nucleus, which is $^{17}$F here.

The initial wave function $\Psi_i$ of the system [$^{17}$F + ${\rm p}]^{J_i^\pi}$ is:
\begin{eqnarray}
\label{wfi}
\Psi_i(r) = \sum_{l_a,j_a,I_t} i^{l_a} 
              {\Psi^{J_i}_{l_a,j_a,I_t}(r) \over {k_ar}}
	      \left[(Y^{l_a} \otimes \chi^{s})^{j_a} 
	              \otimes \chi^{I_t} \right]^{J_i}_{m_i}	      
\end{eqnarray}
where $\Psi^{J_i}_{l_a,j_a,I_t}(r)$ describes the radial part of the wave function (\ref{wf})
with one particle in the continuum. The total angular
momentum in the initial channel, $J_i$, is the result of coupling the
target spin $I_t$ with the total angular momentum $j_a$ being carried by the
projectile which itself is obtained by coupling the intrinsic spin $s$ of the
projectile and its relative orbital angular momentum $l_a$. 

The final wave function $\Psi_f$ for the [$^{18}$Ne]$^{J_f^\pi}$ coupled to
the final state total angular momentum, $J_f$, is:
\begin{eqnarray}
\label{wff}
\Psi_f(r) = \sum_{l_b,j_b,I_{t_b}} A^{j_bI_{t_b}J_f}_{l_bsj_b}            
              {u^{J_f}_{l_b,j_b,I_{t_b}}(r) \over {r}}
	      \left[(Y^{l_b} \otimes \chi^{s})^{j_b} 
	         \otimes \chi^{I_{t_b}} \right]^{J_f}_{m_f}
\end{eqnarray}
where
$A^{j_bI_{t_b}J_f}_{l_bsj_b}$ is the coefficient of fractional parentage and 
$u^{J_f}_{l_b,j_b,I_{t_b}}(r)$ is the s.p. wave in the 
many-body state $J_f$. $s, l_b$ and $j_b$ represent the spin, the orbital
angular momentum and the total angular momentum of the captured
 nucleon (projectile), respectively and $I_{t_b}$ is the final state target spin.
The explicit summations over the target spin state in Eqs. (\ref{wfi}) and (\ref{wff}) 
ensures that we take into consideration 
the capture from the g.s. as well as the excited states of the
target nuclei.

With the wave functions  $\Psi_i(r)$ and $\Psi_f(r)$, we can calculate the electric 
transition amplitudes of multipolarity $\lambda$ as:
\begin{eqnarray}
\label{te}
T^{E\lambda} &=& C(E\lambda) i^{l_a} {\hat J}_f {\hat l}_b {\hat j}_b {\hat j}_a
           \langle \lambda \theta J_f m_f| J_i m_i\rangle  
	   \langle l_b 0 \lambda 0 |l_a 0 \rangle \nonumber \\
	  & & \times W(j_b I_t \lambda J_i; J_f j_a) W(l_b s \lambda j_a;j_bl_a)
	      \delta_{I_tI_{t_b}}I^{\lambda,J_i}_{l_aj_a,l_bj_b}  
\end{eqnarray}
and the magnetic dipole transition as:
\begin{eqnarray}
\label{tm}
T^{M1} &=& i^{l_a} \mu_N {\hat J}_f \langle 1 \theta J_f m_f| J_i m_i\rangle 
           \{ W(j_bI_t1J_i;J_fj_a){\hat j}_a{\hat j}_b \nonumber \\
    &&  \times \left[\mu ({Z_t\over m_t^2} + {Z_a\over m_a^2})
        {\hat l}_a {\tilde l}_a W(l_bs1j_a;j_bl_a) 
	+ (-1)^{j_b-j_a} 2 g_a {\hat s} {\tilde s} W(sl_b1j_a;j_bs) \right] \nonumber \\
    &&  + g_t 	(-1)^{J_f-J_i}{\hat I}_t {\tilde I}_t W(I_tj_b1J_i;J_fI_t) \delta_{j_aj_b}\} 
         \delta_{l_al_b}\delta_{I_tI_{t_b}} I^{0,J_i}_{l_aj_a,l_bj_b}
\end{eqnarray}
In the above formulae, $\theta = m_i - m_f$, ${\hat a} \equiv \sqrt{(2a+1)}$, 
${\tilde a} \equiv \sqrt{a(2a+1)}$, and
\begin{eqnarray*}
C(E\lambda) = \mu^\lambda \left({Z_a\over m_a^\lambda} (-)^\lambda 
                  {Z_t\over m_t^\lambda}\right)
\end{eqnarray*}
$\mu$ is the reduced mass of the system, $m_a$ and $m_t$ are the masses, 
$Z_a$ and $Z_t$ are the charges and $g_a$ and $g_t$ are the gyromagnetic 
ratios of the projectile and the target, respectively.
$\mu_N$ is the nuclear magneton and $I^{\lambda,J_i}_{l_aj_a,l_bj_b}$, the overlap
integral is given by:
\begin{eqnarray*}
I^{\lambda,J_i}_{l_aj_a,l_bj_b} = 
             \int dr u^{J_f}_{l_b,j_b,I_{t_b}}(r) r^\lambda \Psi^{J_i}_{l_a,j_a,I_t}(r)	     
\end{eqnarray*}
The Kronecker symbol $\delta_{I_tI_{t_b}}$ in eqs. (\ref{te}) and (\ref{tm}), ensures that the double summation over the target spins arising from eqs. (\ref{wfi}) and (\ref{wff}) is reduced to a single one (over $I_t$, say). The radiative capture cross-section is then expressed as:
\begin{eqnarray}
\sigma^{E1,M1} &=& {{16\pi} \over 9} \left({\mu \over \hbar c}\right)
              \left({e^2 \over \hbar c}\right)
             \sum_{m_i m_f} \sum_{l_aj_al_bj_bI_t} 
	     {1 \over {\hat s} {\hat I_t}} \left({k_\gamma^3 \over k_a^3}\right) |T^{E1,M1}|^2 \\
\sigma^{E2} &=& {{4\pi} \over 75} \left({\mu \over \hbar c}\right)
              \left({e^2 \over \hbar c}\right)
             \sum_{m_i m_f} \sum_{l_aj_al_bj_bI_t} 
	     {1 \over {\hat s} {\hat I_t}} \left({k_\gamma^5 \over k_a^3}\right) |T^{E2}|^2	     
\end{eqnarray}
where $k_\gamma$ and $k_a$ are the wavevectors of the emitted photon and the
incoming projectile, respectively. These formulae take into account radiative
capture from both the excited state and the g.s. of the target nucleus.

\section{Results and discussion}
In this section we shall present the SMEC results for the spectrum of 
$^{18}$Ne and then go on to discuss the astrophysical S-factor and
the reaction rate for the $^{17}$F(p,$\gamma$)$^{18}$Ne radiative proton capture reaction.

\subsection{The self-consistent potential}
\label{sect_self_const}
The $Q$ subspace in SMEC is constructed by the self-consistent, iterative method
which for a given initial average s.p.\ potential (\ref{pot}) and for a given residual two-body interaction between $Q$ and $P$ subspaces ({\ref{force}) yields the self-consistent s.p.\ potential depending on the
s.p.\ wave function $l_j$, the total angular momentum $J$ of the $A$-nucleon system \cite{karim,phyrep}. The channels in $P$: $(J_{A-1}^{\pi},l_j)^{J}$, are defined by the states $J_{A-1}^{\pi}$ of the $A-1$ system and the s.p. wave function $l_j$; both are coupled to $J$ in the $A$-system. The parameters for the initial potential $U(r)$ used in the calculations for the self-consistent potential $U^{sc}(r)$ are
$V_0=56.8853$ MeV, $V_{SO}=4.9014$ MeV, $R_0=3.276$ fm and $a=0.58$ fm. The potential radius $R_0$ has been defined using the convention $R_0 = r_0(A_t^{1/3})$, where $r_0=1.27$ fm
and $A_t=A-1$ is the mass number of the target. In the self-consistent potential, we obtain
the proton s.p. orbit $1s_{1/2}$ at the experimental binding energy (-2.034 MeV) of the $J^{\pi}=2_{1}^{+}$ level of $^{18}$Ne and the $0d_{5/2}$ s.p. orbit at the experimental binding energy (-3.92 MeV) of the g.s. $J^{\pi}=0_{1}^{+}$  of $^{18}$Ne. This also puts the $0p_{1/2}$ and $0f_{7/2}$ s.p. orbits at  -15.3 MeV and 8.52 MeV, respectively. Another set of parameters which enter into the calculations are those of the density-dependent residual interaction ({\ref{force}). These are
determined to be: ${\hat v}_{00}^{in} = 53.6,
{\hat v}_{00}^{ex} = -438.1, {\hat v}_{01}^{in} = 160.9,
{\hat v}_{01}^{ex} = 169.9$ all in units of of MeV$\cdot$fm$^3$, 
$\nu = 1.374$, $r_n = 2.64$ fm and $d=0.58$ fm,
such that the experimental width \cite{bard} 
of the many-body state  $J^{\pi}=3^{+}$ in $^{18}$Ne is reproduced. 

\subsection{Spectrum of $^{18}$Ne}

\begin{figure}
\centering
\includegraphics*[height=15cm,width=13cm]{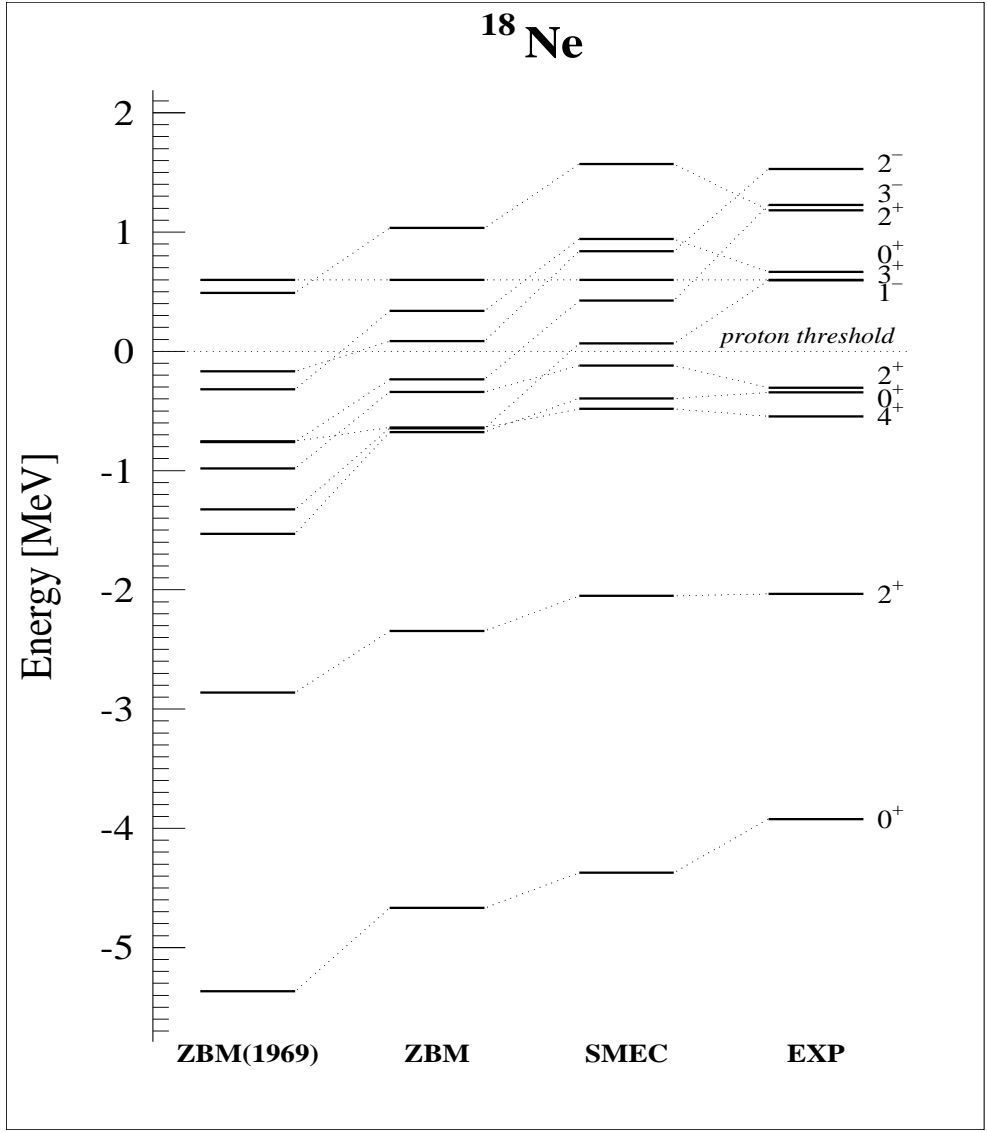}
\caption{\label{spec} The spectrum of $^{18}$Ne. For more details, see the discussion in the text.}
\end{figure}
Fig.\ref{spec} presents the energy spectrum of $^{18}$Ne calculated both
in SM and in SMEC
using the Zuker-Buck-McGrory (ZBM) interaction \cite{zbm} 
in the $Q$ subspace. SM results with the original ZBM force \cite{zbm} are shown
on the l.h.s. of Fig.\ref{spec} and are labelled by 'ZBM(1969)'. With 
certain changes in the matrix elements of the effective interaction (cf Table \ref{tabzbm})
we generate the SM spectrum labelled by 'ZBM'.
This interaction is used also to generate the SMEC spectrum (labelled 'SMEC'). 
The experimental energy levels (labelled 'EXP') are plotted on the r.h.s. of Fig.\ref{spec}. 
The experimental position of the $J^{\pi}=3_1^{+}$ level is taken to be the reference level in the spectra. 
The SM energies , SMEC energy and widths, and also the experimental 
data are presented in Table \ref{tab18ne}. 
\begin{table}
\caption{The antisymmetrized two-body matrix elements in ZBM (this work)
which differ from the original ZBM interaction \cite{zbm}. Here, $d\equiv 0d_{5/2}$ and 
$s\equiv 1s_{1/2}$. }
\vspace{1cm} 
\label{tabzbm}
\begin{center}
\begin{tabular}{| c | c | c | c  | c | c |}
\hline
Interaction & $[dd~dd]$ & $[dd ~dd]$ & $[dd~ dd]$ &$[ds~ ds]$ &$[ds~ ds]$ \\
            & T=1, J=0  & T=1, J=2  & T=1, J=4  & T=1, J=2 &T=1, J=3 \\
\hline
ZBM (this work)& $-2.01$&   $-1.61$ & $-0.48$& $-0.57$& $0.56$ \\
ZBM (1969)& $-2.41$&   $-1.21$ & $-0.08$& $-1.17$& $1.16$ \\
\hline
\end{tabular}
\end{center}
\end{table}

\begin{table}[h]
\caption{SM  energies and SMEC energies and widths vs.\ experimental data for $^{18}$Ne.
The experimental data are from: $a\equiv$ Ref. \cite{till}, 
$b\equiv$ Ref. \cite{bard}, $c\equiv$ Ref. \cite{hahn}. Energies and widths are given in keV.}
\vspace{1cm} 
\label{tab18ne}
\begin{center}
\begin{tabular}{| c| c | c | r r  | c  c|}
\hline
State & ZBM (1969) \cite{zbm}&ZBM& 
\multicolumn{2}{ c |}{SMEC} & \multicolumn{2}{ c |}{Experiment}\\
J$^{\pi}$ & energy & energy & energy & width & energy & width \\
\hline
$ 0^+ $ & $-5366 $ & $ -4668 $ & $ -4372 $ & --- & 
 $ -3921.6 \pm 4.7^a $ & { --- } \\
$ 2^+ $ & $ -2861$ &$ -2345 $ & $ -2050 $ & --- & 
 $-2034.3  \pm 0.2^a $ & { --- } \\
$ 4^+ $ & $ -752$ &$ -639 $ & $ -482 $ & --- & 
 $ -545.4 \pm 0.2^a $ & { --- } \\ 
$ 0^+ $ & $ -1529$ &$ -676 $ & $ -395 $ & --- & 
 $ -345.3 \pm 2.0^a $ & { --- } \\ 
$ 2^+ $ & $ -981$ &$ -340 $ & $ -119 $ & --- & 
 $ -305.2 \pm 0.6^a $ & { --- } \\ 
$ 1^- $ & $-1326 $ &$ -644 $ & $ 65 $ & $\approx 0$ & 
 $ 595 \pm 5^b $ &  $ 0.1 \pm 0.1 $\\ 
$ 3^+ $ &$ 600$ & $ 600 $ & $ 600 $ & 18 & 
 $ 600 \pm 2^b $ & $ 18 \pm 2 $ \\ 
$ 0^+ $ & $-319 $ &$ 338 $ & $ 941 $ & 0.91 & 
 $ 666 \pm 5^b $ & $ 1.0  \pm 1.0 $ \\ 
$ 2^+ $ & $ 490$ &$ 1033 $ & $ 1569 $ & 40.2 & 
 $ 1182 \pm 8^c $ & $ 50.0 \pm 10.0 $ \\ 
$ 3^- $ & $-759 $ &$ -235 $ & $ 425 $ & 0.002 & 
 $ 1229 \pm 8^c  $ & $ \leq 20   $ \\ 
$ 2^- $ & $ -167 $ &$ 84 $ & $ 839 $ & 0.104 & 
 $ 1530 \pm 8^c  $ & $ \leq 20  $ \\ 
\hline
\end{tabular}
\end{center}
\end{table}

\subsubsection{Further effects of the continuum coupling}
\begin{figure}
\centering
\includegraphics*[height=10cm]{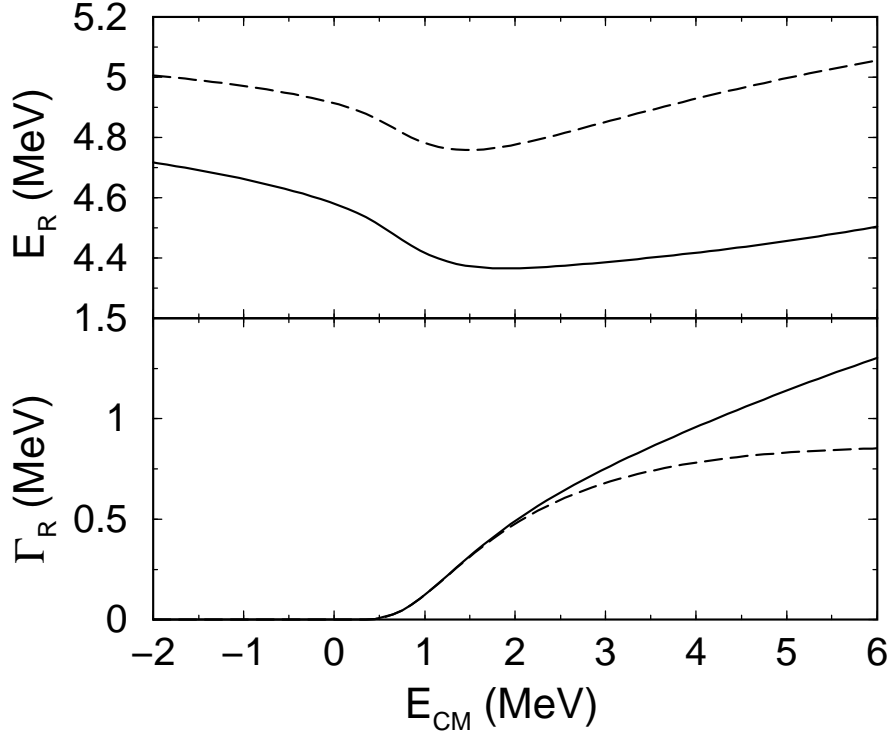}
\caption{\label{bd2} The energy dependence of the eigenvalue (both real $E_R$ 
and imaginary $\Gamma_R$ parts) of the effective Hamiltonian $H_{QQ}^{eff}$
for the $3_1^{+}$ state in $^{18}$Ne. The solid line corresponds to inclusion
of couplings to both the g.s. and the first excited state in $^{17}$F, while the dashed
line corresponds to the coupling to the g.s. of $^{17}$F only.}
\end{figure}
\begin{figure}
\centering
\includegraphics*[height=10cm]{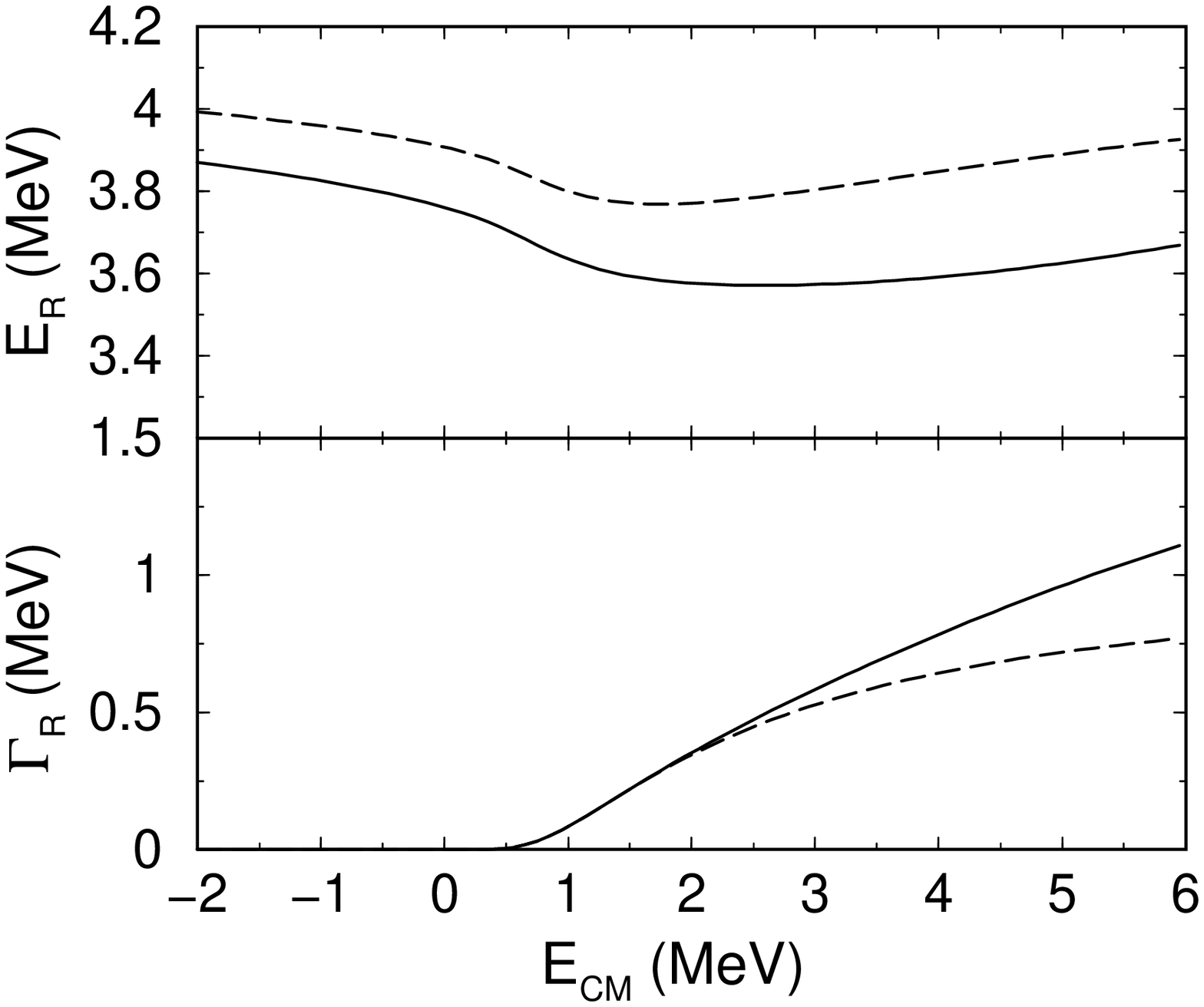}
\caption{\label{bd2_2p} The same as in Fig. \ref{bd2} but
for the $2_{2}^{+}$ in $^{18}$Ne. For more information see the caption of Fig. \ref{bd2} and the discussion in the text.}
\end{figure}
Figs.~\ref{bd2} and \ref{bd2_2p} show the energy dependence of the eigenvalue
of the effective Hamiltonian $H_{QQ}^{eff}$ (eq. (\ref{heff})) for $J^{\pi}=3_1^{+}$
and  $2^{+}_2$, respectively.  The purpose of these figures is
not only to show the energy dependence of the eigenvalues of the effective Hamiltonian
but also to point out that the continuum coupling affects both the bound and the
resonant states in the energy spectra. The top half of the figure, in both cases,
shows the real part ($E_R$) and the bottom half shows the imaginary part ($\Gamma_R$)
of the eigenvalue as a function of the c.m. energy ($ E_{CM}$). 
In all cases the solid line is the result which includes couplings both to the $5/2^+$ g.s. and to the
$1/2^+$ excited state of $^{17}$F, while the dashed line shows the result
which includes couplings from the $5/2^+$ g.s. only. It is interesting to
note (see Fig. ~\ref{bd2}) that $E_R$ differs nearly by 400 keV, at the position
of the $J^{\pi}=3_1^{+}$ resonance, for the calculation with both
the g.s. and the excited state of $^{17}$F as against only with the g.s. This shows
that the coupling to the excited state is essential for the description
of the real part of the $J^{\pi}=3_1^{+}$ energy eigenvalue. The sensitivity of
 the width ($\Gamma_R$) of the $J^{\pi}=3_1^{+}$ state to the coupling 
 to the excited state is rather weak at low excitation energies. At higher energies, this coupling 
 strongly modifies the external mixing of unperturbed SM states,  changing qualitatively the energy dependence of the width (see Fig. \ref{bd2}). 
 
 A similar inference can be drawn from the energy dependence of the 
eigenvalue for the $J^{\pi}=2^{+}_2$ state in $^{18}$Ne (see Fig. ~\ref{bd2_2p}). 
Inclusion of the excited state along with the g.s. of $^{17}$F results in a difference of about 150 keV at the position of the $2^{+}_2$ state with respect to the calculation without the excited state of $^{17}$F. However, considering the relative energy shifts, 
including the coupling to the excited state seems to be more 
important for the  $3_1^{+}$ than for the $2^{+}_2$ state. Similarly as for the $J^{\pi}=3_1^{+}$ state,  the imaginary part of the $J^{\pi}=2_2^{+}$ energy eigenvalue 
 saturates at higher $E_{CM}$ in the absence of the coupling to the excited state of $^{17}$F.

In the following, we shall study of the effect of the continuum coupling on the spectroscopic factors in the 
$J^{\pi}=2^{+}_2$ and $J^{\pi}=3^{+}_1$ eigenvalues of the effective Hamiltonian $H_{QQ}^{eff}$.
 In SMEC, the spectroscopic factor, like the expectation value of any other operator ${\hat O}$, can be calculated as:
\begin{eqnarray}
\langle {\hat O} \rangle = {{{\langle \tilde {\Phi}_j}|{\hat O}| {\tilde {\Phi}_j}\rangle} \over 
                             {{\langle \tilde {\Phi}_j}| {\tilde {\Phi}_j} \rangle}} 	     
\end{eqnarray}
where $\tilde {\Phi}_j$ (cf eq. (\ref{transf})) is the eigenvector of $H_{QQ}^{eff}$.
For the case of the spectroscopic factor one identifies: 
\begin{eqnarray}
{\hat O} = a^\dagger |t \rangle \langle t| a
\end{eqnarray}
where $|t \rangle$ is the target state of the $(A-1)$-system and 
$a^\dagger$ and $a$ are creation and annihilation operators, respectively.

\begin{figure}
\centering
\includegraphics[height=10cm]{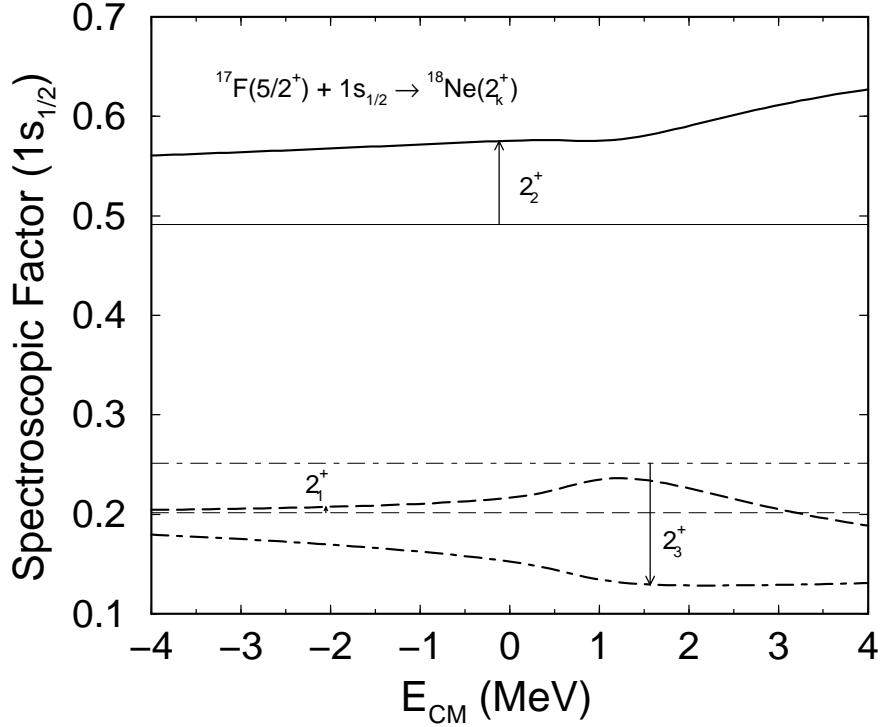}
\caption{\label{spft2gss}$1s_{1/2}$ spectroscopic factors  to the g.s. of $^{17}$F: $[^{17}{\rm F}(5/2_1^+)+1s_{1/2}]\rightarrow {^{18}{\rm Ne}}(2_k^+)$~, $(k=1,2,3)$, for $2^+$ states 
in $^{18}$Ne as calculated in SMEC (thicker lines) and SM (thinner lines). The dashed, solid and dot-dashed lines show it for  $2^+_1$, $2^+_2$ and $2^+_3$ states, respectively. The change of the 
spectroscopic factors from the SM to SMEC values are indicated by 
arrows in each case at the experimental positions of these states (cf Table 2).}
\end{figure}
\begin{figure}
\centering
\includegraphics[height=10cm]{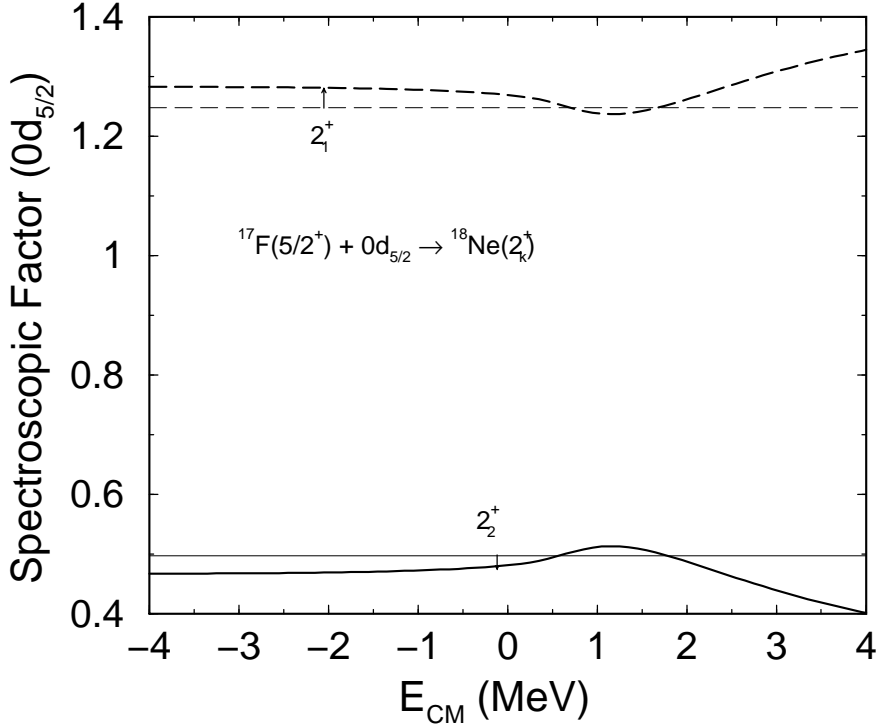}
\caption{\label{sfgsd}$0d_{1/2}$ spectroscopic factors to the g.s. of $^{17}$F: $[^{17}{\rm F}(5/2_1^+)+0d_{5/2}]\rightarrow {^{18}{\rm Ne}}(2_k^+)$~, $(k=1,2)$, for $2^+$ states 
in $^{18}$Ne as calculated in SMEC (thicker lines) and SM (thinner lines) . For more details, see the caption of Fig. \ref{spft2gss} and the discussion in the text.}
\end{figure}
\begin{figure}
\centering
\includegraphics[height=10cm]{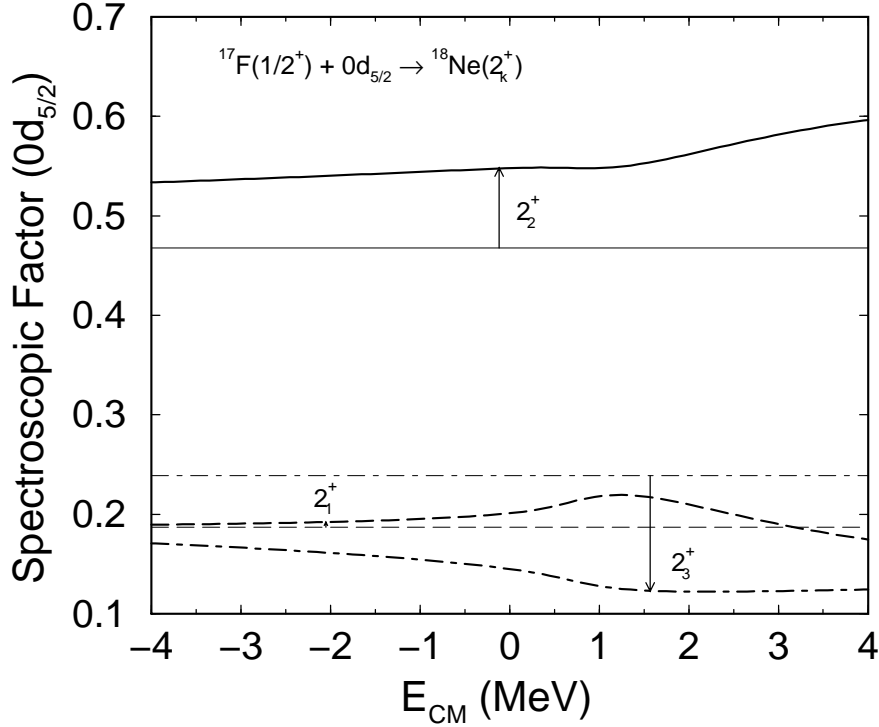}
\caption{\label{sfexd}$0d_{5/2}$ spectroscopic factors to the excited $1/2_1^+$ state in $^{17}$F: $[^{17}{\rm F}(1/2_1^+)+0d_{5/2}]\rightarrow {^{18}{\rm Ne}}(2_k^+)$~, $(k=1,2,3)$  for $2^+$ states in $^{18}$Ne. The dashed, solid and dot-dashed lines show it for  
$2^+_1$, $2^+_2$ and $2^+_3$ states, respectively. SMEC values are plotted with thicker lines. For other details, see the caption of Fig. \ref{spft2gss} and the discussion in the text.}
\end{figure}
\begin{figure}
\centering
\includegraphics[height=10cm]{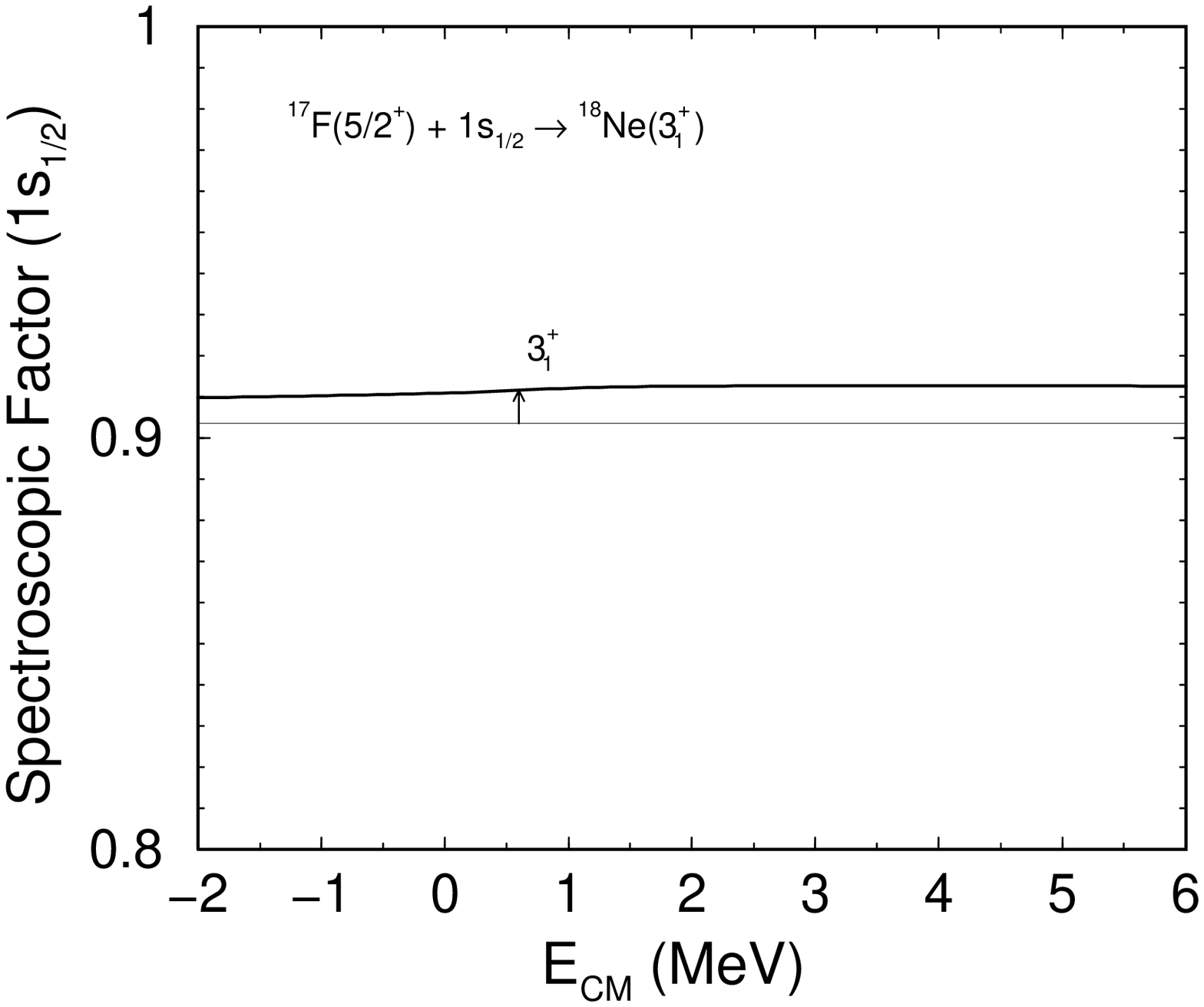}
\caption{\label{spft}$1s_{1/2}$ spectroscopic factor  to the g.s. of $^{17}$F: $[^{17}{\rm F}(5/2_1^+)+1s_{1/2}]\rightarrow {^{18}{\rm Ne}}(3_1^+)$~, for $3_1^+$ state
in $^{18}$Ne as calculated in SMEC (thicker line) and SM (thinner line) . The change of the 
spectroscopic factors from the SM to SMEC values is indicated by an
arrow at the experimental position of this state (cf Table 2).}
\end{figure}

In Fig. \ref{spft2gss}, we show the $1s_{1/2}$ spectroscopic factors $[^{17}{\rm F}(5/2_1^+)+1s_{1/2}]\rightarrow {^{18}{\rm Ne}}(2_k^+)$~, $(k=1,2,3)$
in SMEC (thicker lines) for the first three $2^+$ states in $^{18}$Ne and
compare them with those obtained in SM (thinner lines). 
The SMEC spectroscopic factors are energy dependent, as a consequence of the energy-dependence of $H_{QQ}^{eff}$. Fig. \ref{spft2gss} also shows that the coupling to the continuum
has an overall effect of strongly modifying the $s$-state content of the $2^+_2$ 
and $2^+_3$ states, increasing and decreasing them, respectively. On the contrary, the well-bound $2^+_1$ state is weakly affected by this coupling. This redistribution
of SM spectroscopic factors is a clear indication of the strong mixing of a weakly bound $2_2^+$ state and a resonance $2_3^+$ due to the $Q$-$P$ continuum coupling in the channel $(5/2_1^+,s_{1/2})^{2^+}$. One may also notice that the external mixing of $2^+$ SM states changes at higher energies close to the position of the $2_3^+$ resonance $(E_r=1.57$ MeV). 

$0d_{5/2}$ spectroscopic factors $[^{17}{\rm F}(5/2_1^+)+0d_{5/2}]\rightarrow {^{18}{\rm Ne}}(2_k^+)$~, $(k=1,2)$ in SMEC and SM for the first two $2^+$ states in $^{18}$Ne are shown in Fig. \ref{sfgsd}. The change of the spectroscopic factors by the external mixing is here less important and involves mainly  $2^+_1$ and $2^+_2$ states. For the $2^+_3$ state, the spectroscopic factors in SM and SMEC are both very similar and small. 

$0d_{5/2}$ spectroscopic factors $[^{17}{\rm F}(1/2_1^+)+0d_{5/2}]\rightarrow {^{18}{\rm Ne}}(2_k^+)$~, $(k=1,2,3)$ to the excited $1/2_1^+$ state in $^{17}$F are shown in Fig. \ref{sfexd}. The channel $(1/2_1^+,d_{5/2})^{2^+}$ dominates here the external mixing.
The energy dependence of SMEC spectroscopic factors is similar as found for 
$1s_{1/2}$ spectroscopic factors to the g.s. of $^{17}$F (cf Fig. \ref{spft2gss} ). Again, the redistribution of the SM spectroscopic factors among SMEC $2^+$ states concerns mainly a weakly bound $2^+_2$ state and a $2^+_3$ resonance. 

The continuum coupling for $2_2^+$ state, as probed by the energy dependence of its eigenvalue and the redistribution of the spectroscopic factors, goes essentially through the coupling to the channels: 
$(5/2_1^+,s_{1/2})^{2^+}$ and $(1/2_1^+,d_{5/2})^{2^+}$.
 The $\ell=2$ ($d$-state) coupling to the excited $1/2_1^+$ state of $^{17}$F , changes in particular
the weakly bound $2_2^+$ state (cf Fig. \ref{bd2_2p}). This part of the continuum 
 coupling for $2^+$ states will be strongly suppressed in the one-proton capture cross-section 
 which is  sensitive mainly to low-${\ell}$ ($\ell=0,1$) values. The role of the $1/2_1^+$ 
 state of $^{17}$F for the rate of the reaction $^{17}$F(p,$\gamma$)$^{18}$Ne will be discussed in sect. 
 {\it 3.4}.

Fig. \ref{spft} shows the $1s_{1/2}$ spectroscopic factor  $[^{17}{\rm F}(5/2_1^+)+1s_{1/2}]\rightarrow {^{18}{\rm Ne}}(3_1^+)$~, as calculated in SMEC and SM. The redistribution of the $s$-state spectroscopic factors among different $3^{+}$ SM states by the continuum coupling is here small and almost independent of the total energy of the system.

\subsection{The astrophysical S - factor for the $^{17}${\rm F}(${\rm p},\gamma$)$^{18}{\rm Ne}$ reaction}
We now present the astrophysical S-factor for the capture reaction from both the 
$5/2_1^+$ g.s. and the $1/2_1^+$ excited state of $^{17}$F. In the SMEC calculations, all 
relevant transitions from the initial ($J_i^\pi = 1^-,2^-,3^-,2^+,3^+$) correlated many-body continuum
states in  $^{18}$Ne to the final bound states ($J_f^\pi = 0^+_{1,2},2^+_{1,2},4_1^+$) appropriate for E1,M1 and E2 transitions have been included\footnote{The subscripts $i$ and $f$ in 
$J_i^\pi$ and $J_f^\pi$, respectively, denote the initial and final states
of the transition process.}.
Since the position of the negative parity
states in our calculations are somewhat lower than those experimentally observed,
we have shifted the position of $J_i^\pi = 1^-$ to its experimental value
and have used the same shift for all calculations involving the negative parity states.

In Fig. \ref{sfall}, we present the total S-factor (solid line) for
the $^{17}$F(p,$\gamma$)$^{18}$Ne reaction and the separate contributions: S$_{E1}$ 
for E1 transitions (dotted line), S$_{M1}$ for M1 transitions (dashed line) and S$_{E2}$ for E2 transitions (dot-dashed line). The S$_{E2}$ in Fig. \ref{sfall}, which is orders of magnitude smaller than the 
S$_{E1}$ and S$_{M1}$, is multiplied by 1000. Thus, S$_{E1}$ and S$_{M1}$ components determine
the reaction rate for the proton capture by $^{17}$F. At low excitation energies, S$_{M1}>$S$_{E1}$ and above the $3_1^+$ resonance the S$_{E1}$ dominates. The value of the S-factor at zero energy, $S(0)$, is 2130.29 eV-b and the slope, ${\partial S}/{\partial E_{CM}|_{E_{CM}=0}}$, is $-6.68\times 10^{-3}~\rm b$. For M1 transitions, we find S$_{M1}(0)$=1441.03 eV-b and  ${\partial {\rm S}_{M1}}/{\partial E_{CM}|_{E_{CM}=0}}$=-6.84$\times 10^{-3}$ b, whereas for E1 transitions we have S$_{E1}(0)$=689.26 eV-b and  ${\partial {\rm S}_{E1}}/{\partial E_{CM}|_{E_{CM}=0}}$=+2.13$\times 10^{-4} {\rm b}$.

We will now study the important contributions to the S-factor coming from E1 and M1 transitions. 
The contributions to the total S-factor due to transitions to the different final states of $^{18}$Ne are shown in Fig. \ref{sffs}. Transitions to the $J_f^\pi = 2^+_1$ bound state of $^{18}$Ne 
dominate over all other contributions. This fact is also borne out by calculations presented in Ref. \cite{des}. 
\begin{figure}
\centering
\includegraphics[height=10cm]{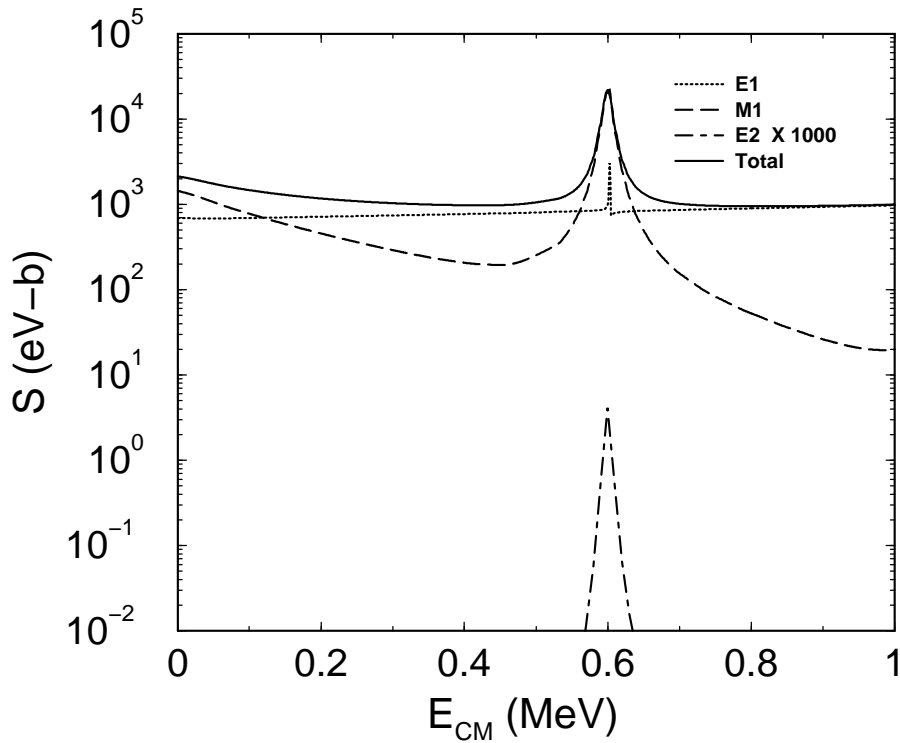}
\caption{\label{sfall} S-factor (solid line) for
the $^{17}$F(p,$\gamma$)$^{18}$Ne reaction and the total contributions
from the E1(dotted line), M1(dashed line) and E2(dot-dashed line).}
\end{figure}

\begin{figure}
\centering
\includegraphics[height=10cm]{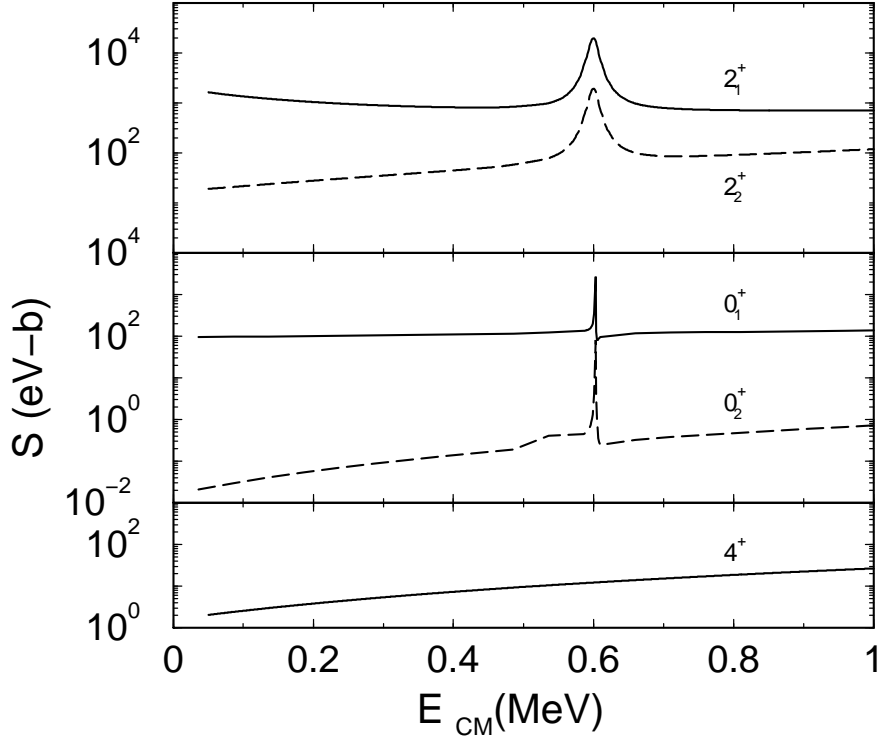}
\caption{\label{sffs}The S-factor in SMEC is shown separately for transitions to different final states of 
$^{18}$Ne. }
\end{figure}

\begin{figure}
\centering
\includegraphics[height=10cm]{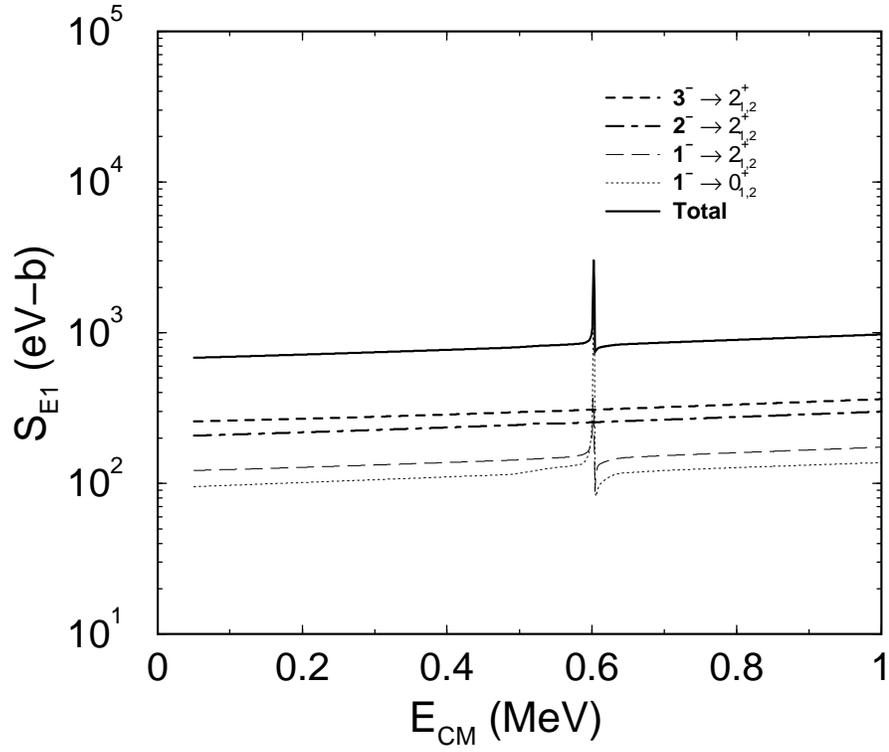}
\caption{\label{e1} E1 part of the S-factor and the individual 
contributions relevant for the $^{17}$F(p,$\gamma$)$^{18}$Ne capture reaction.}
\end{figure}

\begin{figure}
\centering
\includegraphics[height=10cm]{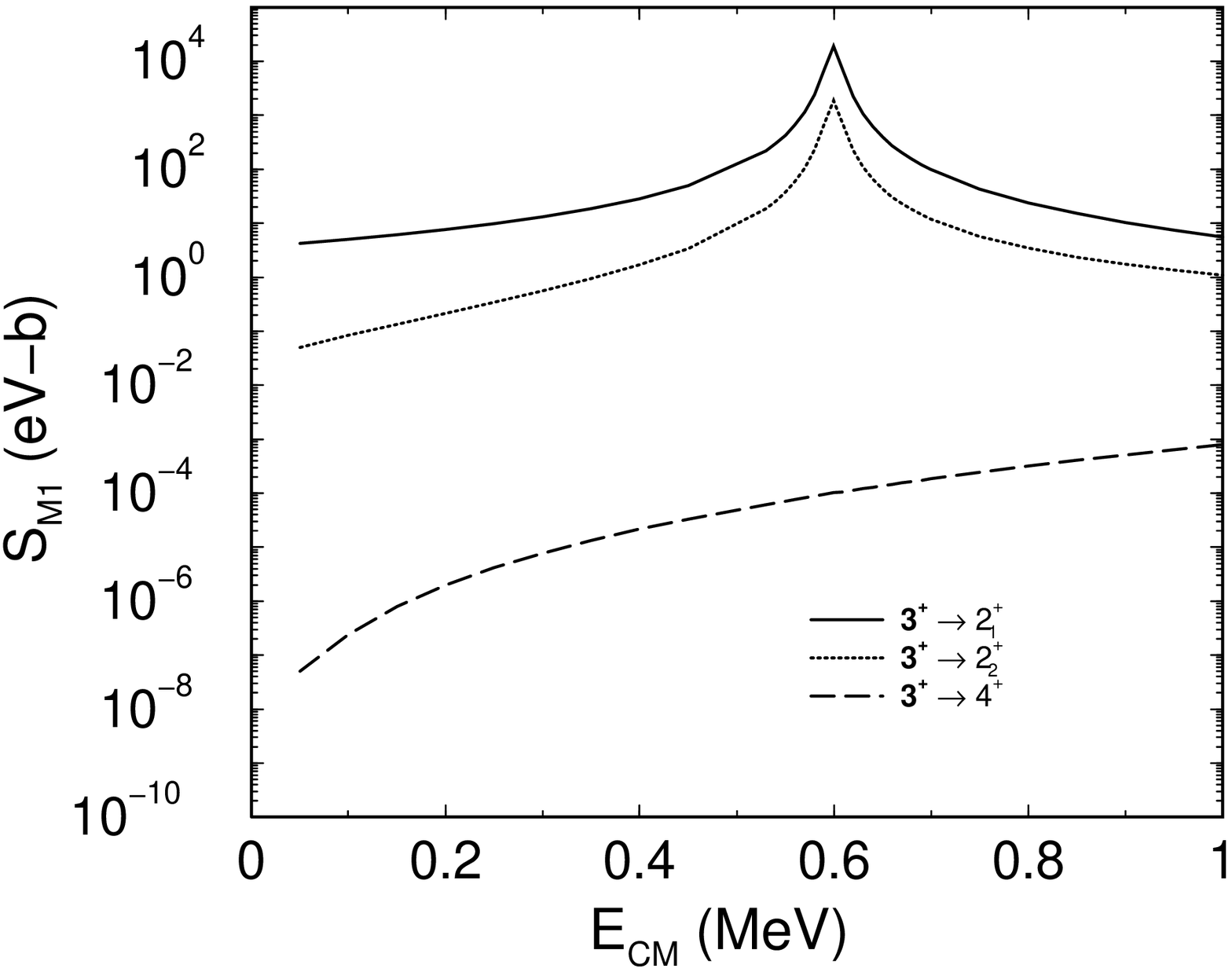}
\caption{\label{m1a}Contributions to S$_{M1}$ from 
$J_i^\pi = 3^+$ to $J_f^\pi = 2^+_{1,2}, 4_1^+$ transitions in $^{18}$Ne.}
\end{figure}

Fig. \ref{e1} shows the S$_{E1}$-factor and the individual contributions relevant for the capture reaction.
Since the experimental $J_i^\pi = 3_1^-,2_1^-$ states in $^{18}$Ne are at energies
above 1 MeV, we show only their non-resonant contribution to the S$_{E1}$-factor.
The non-resonant contribution from the $J_i^\pi = 3^-,2^-$, to  
$J_f^\pi = 2^+_{1,2}$ final states are shown by thick dashed and dot-dashed
lines, respectively.  The contributions from the $J_i^\pi = 1^-$ initial
state to $J_f^\pi = 2^+_{1,2} ~{\rm and}~ 0^+_{1,2} $ are shown by
long dashed and dotted lines, respectively. The total S$_{E1}$ is represented by the solid line.
In all cases, as has been pointed out earlier, the contributions to the
$J_f^\pi = 2^+_1$ bound state of $^{18}$Ne dominate over all other contributions. 
The low energy direct part of the E1 cross-section in our calculations is less than what has been reported in Refs. \cite{garcia,des} due to a different choice of the potential radius. The physical picture in SMEC  consists of generating the s.p. wave functions of the nucleon (proton) moving
in the potential well created by the nucleons in the target nucleus (i.e., $R_0\sim 17^{1/3}$) and 
{\em not} in the combined well of the target nucleus and the nucleon (i.e., $R_0\sim (17^{1/3}+1^{1/3})$) as used in \cite{rolfs,garcia,des}. This physical picture has been followed in all previous SMEC studies of the radiative capture reactions \cite{karim,nic1,nic2}, giving an excellent description of the E1 component.

Attempts to extract the direct capture cross section of the 
$^{17}$F(p,$\gamma$)$^{18}$Ne reaction from the transfer
reactions measurements are underway \cite{bmon}. This procedure involves measuring
 the $^{14}$N($^{17}$F,$^{18}$Ne)$^{13}$C
proton-transfer reaction cross section  and determining the 
asymptotic normalization coefficients (ANC) for the $^{18}$Ne bound states from a
DWBA analysis of the results. Asuming that the measured
transfer reaction is peripheral, one could then relate the extracted ANC with the direct
capture rate at low energies \cite{xu,gag}. Earlier applications of this method
have been to the $^{16}$O(p,$\gamma$)$^{17}$F \cite{gag} and 
$^{7}$Be(p,$\gamma$)$^{8}$B \cite{xu} reactions to determine the
astrophysical S-factor at low energies. Peripheral transfer reactions would be most sensitive to the tail
of the wave functions beyond the range of the nuclear potential than to the inner
region and hence the strength of this method is supposed to lie in the fact that
the potential model bound state wave function could be well reproduced by a 
linearly scaled Whittaker function in the asymptotic region \cite{mukh}, thereby
reducing the dependence on the potential parameters.
However it is to be mentioned here that the low energy part of the S-factor
may not be entirely due to the pure E1 component for a particular radiative capture
process. Importance of this fact will become more apparent as we continue our 
discussions to the M1 transtions, especially on the 
$J_i^\pi = 2^+$ to $J_f^\pi = 2^+_{1,2}$ M1 transitions.

In Fig. \ref{m1a}, we present the contributions to S$_{M1}$ from 
$J_i^\pi = 3^+$ (continuum states) to $J_f^\pi = 2^+_{1,2}, 4_1^+$ (bound states) transitions in $^{18}$Ne. As remarked earlier, the transitions to the $2^+_1$ final state, shown by the solid line in Fig. \ref{m1a}, dominate peaking at the position of the $3_1^+$ resonance. The transitions to $2^+_2$ and to $4_1^+$ are shown by dotted and dashed lines, respectively. The proton capture from the excited state of $^{17}$F has a negligible effect on the value of S$_{M1}$.
This, at first instance, may sound contradictory to the inferences
from Fig. \ref{bd2}, where inclusion of the excited state couplings were shown
to affect the energy eigenvalues. However, these couplings involve a $d$-state whose contribution to the proton capture is strongly reduced in comparison with the $s$-state contribution. The latter one is present in the radiative capture in the $5/2^+$ (g.s.) channel. Moreover,
the $1/2^+$ excited channel opens at 495 keV, reducing the available energy for the proton capture and, hence, reducing importance of this channel in the proton capture cross-section, in particular at low $E_{CM}$. 

Much more important contribution to S$_{M1}$ comes from transitions from  the
$J_i^\pi = 2^+$ continuum  to the $J_f^\pi = 2^+_{1,2}$ bound states of $^{18}$Ne (see 
Fig. \ref{m1b}). The solid line and the dashed line show the transition
to the $ 2^+_{1}$ and $ 2^+_{2}$ bound states of $^{18}$Ne, respectively. 
The $2^+$ continuum at low excitation energies is strongly correlated by the 
proximity of a weakly bound $2_2^+$ state which induces a resonant-like feature
('resonant-halo') in the $2^+$ continuum  owing to its large $s$-spectroscopic factor 
(cf a discussion of Fig. \ref{spft2gss}). Hence, the dominant M1 component at low
 energies (below 0.4 MeV) comes from the transitions between $2^+_2$-parented continuum states and the well-bound $2_1^+$ state.  The weakly bound $2_2^+$ state plays in this scenario a role of the catalyzer of the proton  capture reaction.
 At higher energies, especially in the vicinity of the $3^+$ resonance at 600 keV we see the dominance
 of $J_i^\pi = 3^+$ to $J_f^\pi = 2^+_{1}$ M1 transitions.
\begin{figure}
\centering
\includegraphics[height=10cm]{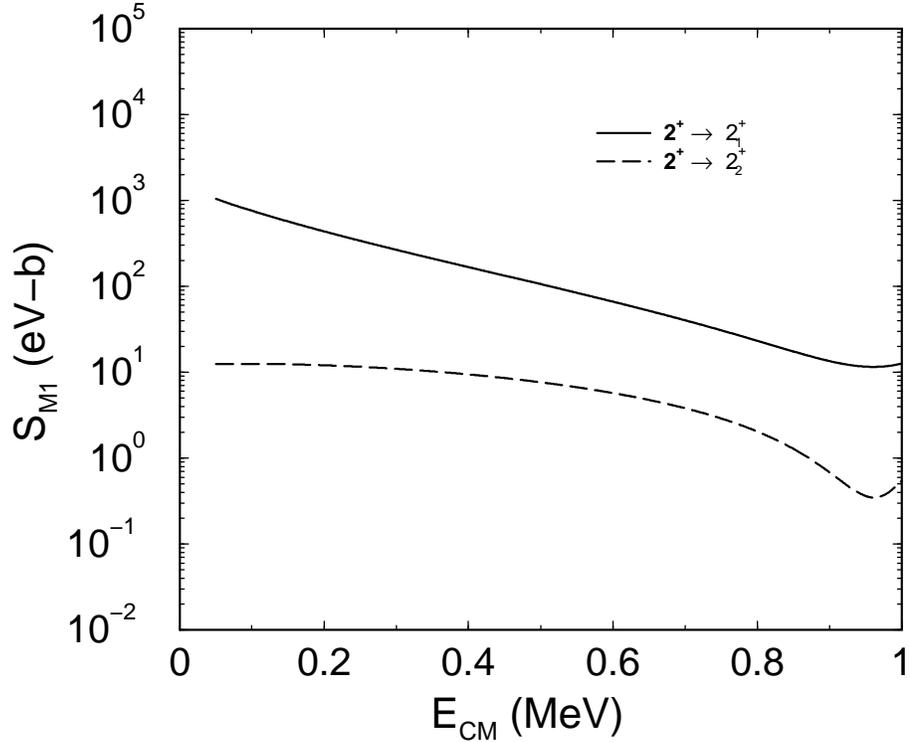}
\caption{\label{m1b}The M1 transitions from 
$J_i^\pi = 2^+$ to $J_f^\pi = 2^+_{1,2}$ bound states of $^{18}$Ne.}
\end{figure}

\begin{figure}
\centering
\includegraphics[height=10cm]{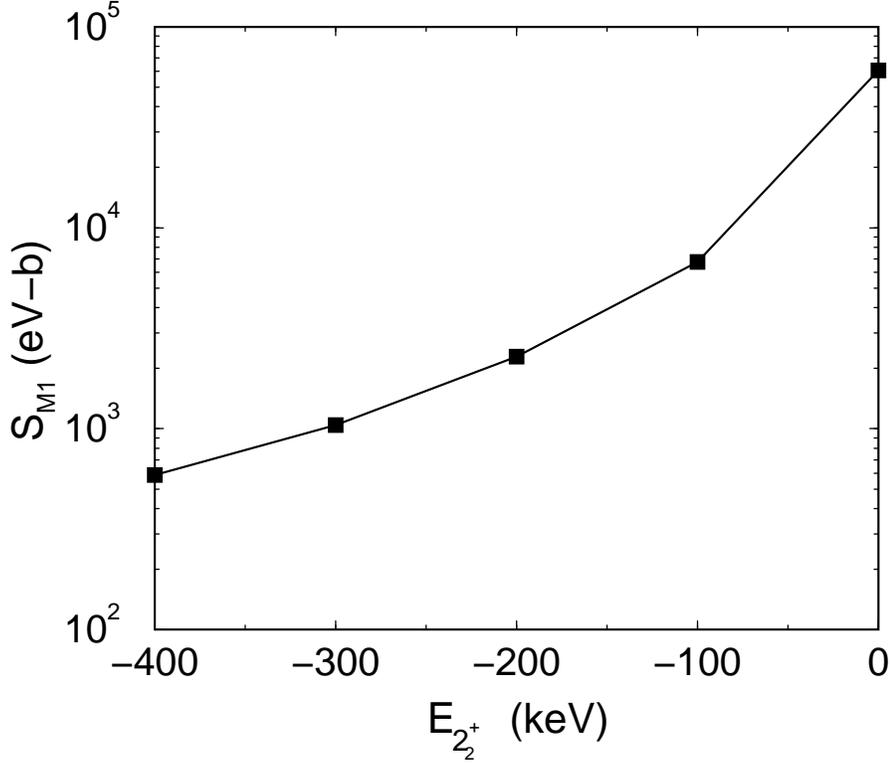}
\caption{\label{th}The S-factor for the 
$J_i^\pi = 2^+$ to $J_f^\pi = 2^+_1$ M1 transition at a fixed energy
$\rm E_{CM} = 50~ keV$, as a function of the position ($\rm E_{2^+_2}$) of the 
weakly bound $2^+_2$ state in $^{18}$Ne with respect to the proton emission
threshold.}
\end{figure}

For a given strength of the continuum coupling ($Q-P$ coupling) and fixed radial s.p. 
wave functions involved in the calculation of the source term \cite{karim,phyrep} 
(cf eq. (\ref{coup})), the rate of the proton capture in M1 transitions: $J_i^{\pi}=2^+\longrightarrow J_f^{\pi}=2_1^+$, depends solely on the position of the $2_2^+$ state with respect to the one-proton emission threshold. In Fig. \ref{th}, we show the S$_{M1}$-factor  at a fixed c.m. energy $E_{CM} = 50$  keV for the $J_i^\pi = 2^+\longrightarrow J_f^\pi = 2^+_1$ transitions, as a function of the energy ($\rm E_{2^+_2}$) of the $2^+_2$ state with respect to the one-proton emission threshold. The WS potential generating the s.p. wave functions is kept fixed in these calculations (see sect. \ref{sect_self_const}). 
One can see from Fig. \ref{th} that changing the energy $E_{2_2^+}$ of the $2_2^+$ state from -400 keV to 0 (the one-proton emission threshold) changes the S$_{M1}$-factor by two orders of magnitude.

This astonishing effect is not due to an increased radial size of the s.p. orbits, in particular of the $1s$-state, involved in the $2_2^+$ state since the average potential is kept fixed for all values of $\rm E_{2^+_2}$. The effect seen in Fig. \ref{th} is a consequence of the strong interference between $2^+_2$ and $2_3^+$ eigenvalues of the effective Hamiltonian which shifts a significant part of the $2_3^+$ resonance strength into the region of low energy  continuum. This is a genuine interference effect in the OQS: states having a similar structure and coupled strongly to the environment of decay channels change in such a way that one state close to the emission threshold aligns its wave function with the decay channel, {\it i.e.} its wave function becomes more similar to the channel wave function \cite{phyrep,iskra,dopr}. A similar segregation effect 
has been found in Refs. \cite{iskra,dopr} in the distribution of widths of the resonance states. Mixing of resonances via the coupling to the decay channels leads to the so-called resonance trapping, {\it i.e.} a major part of the decay width becomes concentrated in one (or more) resonance(s) in the case of one (or more) open channel(s) if the strength of the external coupling is sufficiently strong, whereas remaining states have negligible width and, hence, are trapped. In the considered example of $2^+$ discrete states of $^{18}$Ne, the $1s_{1/2}$ spectroscopic amplitudes in SM states $2_2^+$ and $2_3^+$ are large, indicating a strong mutual coupling (strong external mixing) of these two many-body states via the $\ell=0$ one-particle continuum.  This kind of coupling may lead to the situation that one of the $2^+$ eigenstates of the effective Hamiltonian closest to the one-proton emission threshold ($2_2^+$ in the studied case)   increases its $s$-state content and aligns with the decay channel $[(A-1)\otimes {\rm p}]=[5/2^+\otimes s_{1/2}]$ at the expense of other $2^+$ eigenstates. The proton capture in the $s$-wave then becomes enhanced because the correlations in low-energy continuum states generated by an aligned subthreshold state become enhanced by a downwards shift of the $s$-state strength from the region of $2_3^+$ resonance.
This message was also conveyed by the conclusions drawn from Fig. \ref{spft2gss} which shows that the
coupling to the continuum has the overall effect of strongly modifying 
the $s$-state content of the $2^+_2$ and $2^+_3$ states, and thereby redistributing their strengths.

\begin{table}
\caption{The $1s_{1/2}$ spectroscopic amplitudes in the $2^+_2$ state of $^{18}$Ne for different SM effective interactions.}
\vspace{1cm} 
\label{tabsa}
\begin{center}
\begin{tabular}{| c | c | c |}
\hline
 ZBM (this work) & WBT \cite{wbt} & PSDFP \cite{jim1} \\
 \hline
 0.701 & 0.764 & 0.762 \\
\hline
\end{tabular}
\end{center}
\end{table}

 The effect is all the more accentuated by the large $s$-state spectroscopic amplitude
in the $2_2^+$ state. To verify that this 
is not an artifact of the ZBM interaction, we have also shown in Table \ref{tabsa} the corresponding numbers for two different SM effective interactions, WBT \cite{wbt} and PSDFP \cite{jim1}, also used for  
A = 18 mass region.  We see that the magnitudes of the $s$-state
spectroscopic amplitudes are similar for  all these interactions.

\subsection{The $^{17}{\rm F}({\rm p},\gamma)^{18}{\rm Ne}$ reaction rates}
As has been mentioned earlier the radiative proton capture by $^{17}$F
plays an important role in the nucleosynthesis. In 
Table \ref{tabrate}, we present the reaction rate, $N_A\langle\sigma v\rangle$
in $\rm cm^3 mole^{-1}s^{-1}$ for this reaction 
as a function of some typical stellar temperatures $T$ (in GK).
$\langle\sigma v\rangle$ is the reaction rate per particle pair \cite{rolfsbook} 
and $N_A$ is Avogadro's number. Column 1 of Table \ref{tabrate} shows the SMEC results 
calculated taking into consideration the capture from both the $5/2^+$
g.s. and the $1/2^+$ excited state of $^{17}$F, while the column
designated by SMEC (g.s.) shows the result for proton capture from the 
$5/2^+$ g.s. of $^{17}$F alone. We do not see any noticeable 
difference in the reaction rates in these two calculations, particularly
below temperatures of 0.4 GK which are more important for nova nucleosynthesis.
This is understandable because the Gamow window spans lower energies at these
temperatures and, moreover, the $1/2^+$ excited state channel involves the proton capture in the $d$-wave and is strongly disfavored by the centrifugal barrier.

The last column in Table \ref{tabrate} is the reaction rate from 
Ref. \cite{bard}, which is quoted here for the purpose
of comparison with our reaction rate. The SMEC rates are lower by a
factor of 2-3 below 0.4 GK and by a factor of 7.5 at 2.0 GK, as 
compared with those of Ref. \cite{bard},
primarily because of the lower E1 contributions due to the reasons
mentioned in the previous section.

In Fig. \ref{figrate}, we show the variation of the total reaction rate as a function of 
temperature and for some important transitions.
The total contribution is shown by
the solid line, while the dotted and short-dashed lines represent the
contributions from the E1 and M1 components, respectively. Amongst
the M1 components the contributions from the $J_i^\pi = 2^+$ to 
$J_f^\pi = 2^+_{1,2}$ components (long-dashed) dominate the rate 
at lower temperatures and accounts for almost all the M1 cross-section below
about 0.4 GK, while the contributions from $J_i^\pi = 3^+$ to $J_f^\pi = 2^+_{1,2}$
(dot-dashed) are dominant at higher temperatures. 

\begin{table}
\caption{
$^{17}$F(p,$\gamma$)$^{18}$Ne reaction rates in $\rm cm^3 mole^{-1}s^{-1}$ in
SMEC and Ref.\cite{bard} for some typical stellar temperatures (in GK). For
other descriptions see the text.}
\label{tabrate}
\begin{center}
\vspace{1cm}
\begin{tabular}{|c|c|c|c|}
\hline
T($10^9$K) & SMEC & SMEC (g.s.) & Ref. \cite{bard} \\
\hline
0.1 & 1.317 $\times 10^{-9}$ &1.317 $\times 10^{-9}$ & (2.68$\pm$0.38) $\times 10^{-9}$ \\
0.2 & 2.209 $\times 10^{-6}$ &2.209 $\times 10^{-6}$ & (5.15$\pm$0.75) $\times 10^{-6}$ \\
0.3 & 7.736 $\times 10^{-5}$ &7.736 $\times 10^{-5}$ & (1.97$\pm$0.29) $\times 10^{-4}$  \\
0.4 & 7.458 $\times 10^{-4}$ &7.453 $\times 10^{-4}$ & (2.29$\pm$0.40) $\times 10^{-3}$ \\
0.5 & 4.124 $\times 10^{-3}$ &4.116 $\times 10^{-3}$ & (1.77$\pm$0.49) $\times 10^{-2}$  \\
0.6 & 1.646 $\times 10^{-2}$ &1.640 $\times 10^{-2}$ & (9.29$\pm$3.28) $\times 10^{-1}$	\\
0.7 & 5.023 $\times 10^{-2}$ &5.000 $\times 10^{-2}$ & (3.32$\pm$1.30) $\times 10^{-1}$ \\
0.8 & 1.227 $\times 10^{-1}$ &1.221 $\times 10^{-1}$ & (8.80$\pm$3.61) $\times 10^{-1}$ \\
0.9 & 2.516 $\times 10^{-1}$ &2.503 $\times 10^{-1}$ & (1.88$\pm$0.78) $\times 10^{0}$  \\
1.0 & 4.516 $\times 10^{-1}$ &4.491 $\times 10^{-1}$ & (3.43$\pm$1.44) $\times 10^{0}$ \\
1.5 & 2.643 $\times 10^{0} $ &2.629 $\times 10^{0} $ & (1.97$\pm$0.78) $\times 10^{1}$  \\
2.0 & 6.185 $\times 10^{0} $ &6.155 $\times 10^{0} $ & (4.62$\pm$1.64) $\times 10^{1}$   \\
\hline
\end{tabular}
\end{center}
\end{table}
\begin{figure}
\centering
\includegraphics[height=10cm]{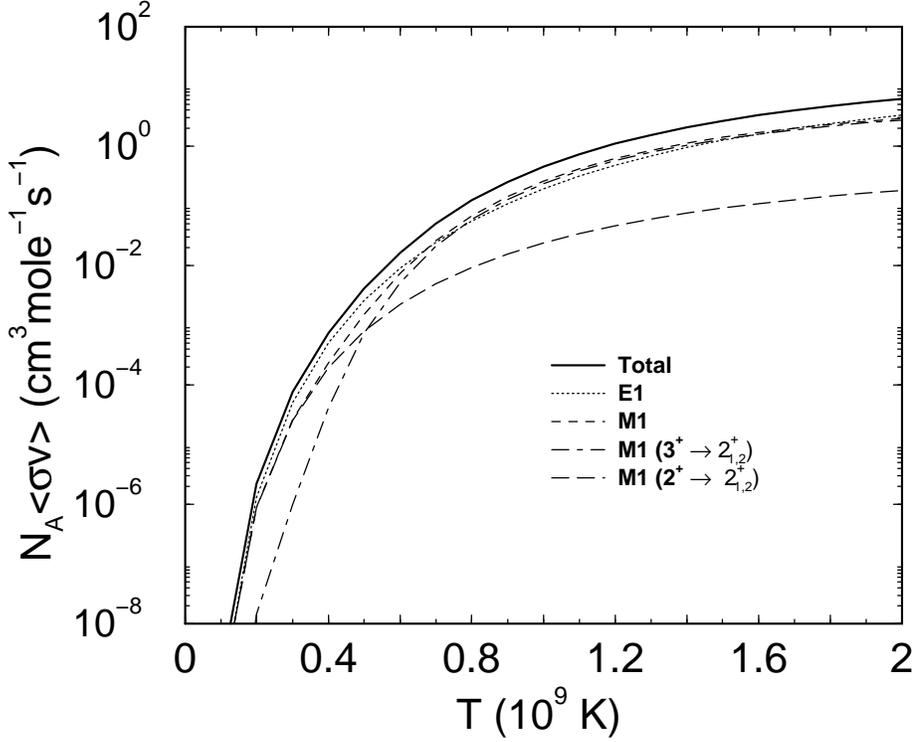}
\caption{\label{figrate}The variation of the total reaction rate as a function of 
temperature and for some important transitions.
The total reaction cross-section is shown by
the solid line, while the dotted and short-dashed lines represent the E1 and M1 components, respectively. For more details, see the description in the text.}
\end{figure}

It would be interesting to investigate the astrophysical implications 
of our $^{17}$F(p,$\gamma$)$^{18}$Ne reaction rate, especially in nova explosions.
The previous reaction rates of Wiescher {\it et al.} \cite{wish} are higher
than that of Bardayan {\it et al.} \cite{bard}
based on the recent determination of the energy and the total width of the
$J^\pi = 3^+$  state in $^{18}$Ne and adding the non-resonant direct capture
rate from the potential model in Ref. \cite{garcia}. The reaction rates 
of Sherr and Fortune \cite{sherr} and Garc{\'i}a {\it et al.} \cite{garcia} 
are rather similar to those of Bardayan {\it et al.} below about 0.4 GK and
are a bit lower at higher temperatures.
The reaction rate of Ref. \cite{bard} was used to calculate \cite{koon} the 
nucleosynthesis in nova outbursts on the surfaces of 1.25 M{$_\odot$} and 
1.35 M{$_\odot$} ONeMg white dwarfs and a 1.00 M{$_\odot$} CO white dwarf.
It was shown that for a  1.25 M{$_\odot$} white dwarf a slower 
$^{17}$F(p,$\gamma$)$^{18}$Ne reaction rate produced more $^{18}$F than 
the faster rate because the $^{18}$F (produced from the $\beta$-decay of 
$^{18}$Ne) at lower temperatures would survive as $^{18}$F and would not be
destroyed by the $^{18}$F(p,$\alpha$)$^{15}$O reaction. 
Although a faster $^{17}$F(p,$\gamma$)$^{18}$Ne reaction rate does allow
more conversion of $^{17}$F to $^{18}$F, but since this 
 $^{18}$F would be produced at higher temperatures it would be destroyed by the
$^{18}$F(p,$\alpha$)$^{15}$O reaction. Thus a larger final abundance of
$^{18}$F would be obtained with a slower $^{17}$F(p,$\gamma$)$^{18}$Ne reaction rate.
 It was also calculated that for a 1.35 M{$_\odot$} white dwarf the abundances of
$^{17}$O and  $^{17}$F increased by a factor of 14000, as against 
estimates based on rates from Ref. \cite{wish},
 in the third hottest zones of the nova. For a 1.00 M{$_\odot$} CO white dwarf,
which involves lower peak temperatures than the other two cases, only small 
variation in the mass fraction of various isotopes were observed except for 
$^{18}$Ne; where it increased by $21\%$ as compared to Ref. \cite{wish}.

Our reaction rate is lower than those of Ref. \cite{bard} by nearly a factor of 2-3
below 0.4 GK. This is also about the temperature range of the hottest zone in 
1.25 M{$_\odot$} and 1.35 M{$_\odot$} white dwarfs, which are 0.333 GK and 
0.457 GK, respectively \cite{koon}.
For a  1.25 M{$_\odot$} white dwarf, with our lower rate, we
would also qualitatively expect a large $^{18}$F abundance produced at
lower temperatures and not being destroyed by the $^{18}$F(p,$\alpha$)$^{15}$O 
reaction due to the same reasons as mentioned in the previous paragraph. 
The final abundances of $^{18}$F, in 1.35 M{$_\odot$} white dwarf, could be
slightly less than in Refs. \cite{wish,koon}, following the same trends as in
Ref. \cite{koon}.
Moreover, recent measurements \cite{kozub} seems to suggest that the estimated
$^{18}$F(p,$\alpha$)$^{15}$O reaction rate could be 2-5 times less than what was
estimated in recent calculations \cite{coc,bard02,ser}. This would imply the 
survival of $^{18}$F ($T_{1/2}=110$ min) and this opens up the possibility 
of detecting the characteristic 511 keV $\gamma$-ray, from the electron-positron 
annihilation following the $\beta$-decay of $^{18}$F, using the techniques of 
gamma-ray astronomy. This, in turn would provide us an opportunity to gain 
further insights about the nova mechanism. 

Furthermore, our lower 
$^{17}$F(p,$\gamma$)$^{18}$Ne reaction rate would allow more $^{17}$F
to survive and the production of more $^{17}$O by $^{17}$F($\beta^+,\nu$)$^{17}$O in both 
1.25 M{$_\odot$} and 1.35 M{$_\odot$} white dwarfs as compared to the values reported in \cite{koon}. Moreover, apart from the branch which produces $^{18}$F via the $^{17}$O(p,$\gamma$)$^{18}$F reaction, the synthesis of $^{15}$O by the 
$^{17}$O(${\rm p},\alpha$)$^{14}$N(${\rm p},\gamma$)$^{15}$O reaction chain would be more feasible with the rates reported in this work. The $\beta$-decay
of $^{15}$O to $^{15 }$N could be a way of explaining the overabundance of $^{15 }$N
in the nova ejecta \cite{chin}.

For a 1.00 M{$_\odot$} CO white dwarf, which involves lower peak temperatures,
we would generally concur with the scenario predicted in Ref. \cite{koon} as our
rates would be closer to the rates of Ref. \cite{bard} at lower temperatures.
We would probably have similar abundances for most of the isotopes as in
Refs. \cite{wish,koon} except for abundances of $^{18}$Ne, which could be 
higher in our case following the similar trends as in Ref. \cite{koon}.

\section{Summary and conclusions}
In this work, we have applied the open quantum system formalism of the SMEC for the microscopic description of the $^{18}$Ne spectra, low energy astrophysical S-factor and reaction rates
for the radiative proton capture reaction $^{17}$F(p,$\gamma$)$^{18}$Ne. 
SMEC is a model in which the realistic SM solutions for (quasi-)bound states are coupled 
to the environment of one-particle \cite{karim,phyrep} and two-particle decay channels \cite{rop} 
for the description of complicated low energy excitations of weakly bound nuclei and various nuclear reactions and decays involving up to two nucleons in the scattering continuum. This 
theoretical model, which attempts a unified description of interdependent nuclear
structure and reaction theories by using a projection operator technique \cite{Fesh}, is a recent development of the Continuum Shell Model (CSM) \cite{csm1,csm2,csm3}. 
 
Inclusion of the target excited state was shown to have an influence on the
eigenvalues of the effective Hamiltonian. It was seen that the energy of the
$3^+$ resonance in $^{18}$Ne could be different by as much as 400 keV, depending
on whether the excited state couplings were included or not. It was also interesting
to observe the energy dependence of the spectroscopic factors for the first three
$2^+$ states in $^{18}$Ne.
The coupling to the continuum strongly modifies the $s$-state content of
the weakly bound $2^+_2$ and resonant $2^+_3$, decreasing the latter 
and reinforcing the former, with respect to their SM values. On the other hand
the $s$-state content of the $2^+_1$ state was not affected too much by
the continuum couplings by virtue of it being well below the particle emission threshold.
It would ve very nice if, in some future experiment, these spectroscopic factors could be 
measured, especially those of the $2^+_1$ and $2^+_2$.

We also calculated the astrophysical S-factors for the radiative capture reaction 
$^{17}$F(p,$\gamma$)$^{18}$Ne, where the capture could occur from both the 
$5/2^+$ g.s. and the $1/2^+$ first excited state of $^{17}$F.
Fully antisymmetrized wave functions were used in both the initial and final
states of the radiative capture process. This method is also fully symmetric in treating the
resonant and non-resonant part of the reaction and one does not need to calculate them separately. 
We have shown that in the region around 600 keV the $J_i^\pi = 3^+$ to $J_f^\pi = 2^+_{1}$ 
component of the M1 dominates the S-factor, but at 
very low energies (below 0.4 MeV) the M1 component arising
from the $J_i^\pi = 2^+$ to $J_f^\pi = 2^+_{1}$ transition has a substantial impact
on the low energy S-factor. In fact its contribution is 
even more than the E1 component below 0.1 MeV. This has been traced back to the strong
mixing of the  weakly bound $2^+_2$ state with the $2^+_3$ resonance
in the low energy correlated $2^+$ continuum wave function. One should, thus
be careful while analyzing the low energy direct capture part of the reaction
with theories, which tend to minimize the effect of nuclear structure in nuclear
reactions. It would indeed be extremely interesting if in proposed experiments
\cite{newex}, the low energy E1 and M1 components could be disentangled and the
low energy behaviour of the M1 is studied in more details.

We have also calculated the reaction rate for the $^{17}$F(p,$\gamma$)$^{18}$Ne
radiative capture reaction for some typical stellar temperatures. Our rates are 
lower by a factor of 2-3 below 0.4 GK and by a factor of 7.5 at 2.0 GK, as 
compared with those of Ref. \cite{bard}. Qualitatively we predict a similar
situation in novae nucleosynthesis, following the trends reported
in Ref. \cite{koon}, with perhaps a larger abundance of $^{17}$F and $^{17}$O.
It would also be interesting to investigate if our smaller reaction rates
could eventually lead to a larger $^{15}$O abundance whose $\beta$-decay
to $^{15 }$N could be a way of explaining the overabundance of $^{15}$N
in the nova ejecta \cite{chin}.


\end{document}